\pdfoutput=1
\documentclass[final,5p,times,twocolumn,numbers,sort&compress,a4paper]{elsarticle}

\usepackage[utf8]{inputenc}
\usepackage{orcidlink}

\usepackage[english]{babel}
\usepackage{amsmath}
\usepackage{amsfonts}
\usepackage{amssymb}
\usepackage{graphicx}
\usepackage{xcolor}
\usepackage{hyperref}
\usepackage{epstopdf}
\usepackage{pdfpages}
\usepackage{placeins} 
\usepackage{booktabs}
\usepackage[super]{nth}
\usepackage[normalem]{ulem} 
\usepackage{xspace} 
\usepackage{lineno}
\modulolinenumbers[1]
\usepackage{caption}
\usepackage{subcaption}
\usepackage{mathtools} 
\usepackage{pdflscape}
\usepackage{breakurl}
\usepackage{multirow}
\usepackage{makecell}

\definecolor{amethyst}{rgb}{0.6, 0.4, 0.8}
\definecolor{green}{rgb}{0.55, 0.71, 0.0}
\definecolor{apricot}{rgb}{0.98, 0.81, 0.69}
\definecolor{auburn}{rgb}{0.43, 0.21,0.1}
\definecolor{babyblueeyes}{rgb}{0.63, 0.79, 0.95}
\definecolor{bittersweet}{rgb}{1.0, 0.44, 0.37}
\definecolor{blue(munsell)}{rgb}{0.0, 0.5, 0.69}
\definecolor{oceanboatblue}{rgb}{0.0, 0.47, 0.75}
\definecolor{brightmaroon}{rgb}{0.76, 0.13, 0.28}
\definecolor{deepcarminepink}{rgb}{0.94, 0.19, 0.22}

\newcommand\arcdeg{\mbox{$^\circ$}\xspace}%
\newcommand\degr{\arcdeg}%
\newcommand{\degC}{\mbox{$^\circ\mathrm{C}$}\xspace}

\usepackage{dblfloatfix}

\begin{document}

\begin{frontmatter}

\title{Direct Comparison of SiPM and PMT Sensor Performances in a large-size imaging air Cherenkov telescope}

\author[MPI_address,TUM_address]{A. Hahn\corref{mycorrespondingauthor}\orcidlink{0000-0003-0827-5642}}
\ead{ahahn@mpp.mpg.de}

\author[MPI_address]{R. Mirzoyan\orcidlink{0000-0003-0163-7233}}
\ead{razmik@mpp.mpg.de}

\author[MPI_address]{A. Dettlaff}
\ead{todettl@mpp.mpg.de}

\author[MPI_address]{D. Fink}
\ead{fink@mpp.mpg.de}

\author[MPI_address,ICRR_address]{D. Mazin\orcidlink{0000-0002-2010-4005}}
\ead{mazin@icrr.u-tokyo.ac.jp}

\author[MPI_address,ICRR_address]{M. Teshima}
\ead{mteshima@icrr.u-tokyo.ac.jp}
 
\cortext[mycorrespondingauthor]{Corresponding author}

\address[MPI_address]{Max Planck Institute for Physics (Werner-Heisenberg-Institut), Boltzmannstr. 8, 85748 Garching, Germany}
\address[TUM_address]{Physics Department, Technical University Munich, James-Franck-Str. 1, 85748 Garching, Germany}
\address[ICRR_address]{Institute for Cosmic Ray Research, The University of Tokyo, 5-1-5 Kashiwa-no-Ha, Kashiwa City, Chiba, 277-8582, Japan}

\begin{abstract}
The peak photon detection efficiency (PDE) of silicon photomultipliers (SiPMs) can be as good or better than the PDE of photomultiplier tubes (PMTs). There are experiments where the signal is measured in the presence of a strong, steady background light emission. In these, one needs to accurately evaluate the signal-to-noise ratio. Imaging Atmospheric Cherenkov Telescopes (IACTs) observe in the presence of strong noise induced by the light of the night sky.  It is certainly interesting to investigate the SiPM performance under operational conditions of IACTs and to compare it with that of the PMTs.\\
For that purpose, we built a SiPM-based detector module, which was installed in one of the imaging cameras of the two Major Atmospheric Gamma-ray Imaging Cherenkov (MAGIC) telescopes in 2015. The experience gained from that module was used to design the second generation of modules of improved performance. Two such modules were installed in 2017.\\
MAGIC is a system of two IACTs located on the Canary Island of La Palma. The mechanical structure of the MAGIC imaging cameras offers the possibility to install up to 6 additional detector modules of 7 pixels each into the open vertices of the hexagonal-shaped camera. This allows us to directly, without making any assumption, compare the performance of the PMT-based modules with that of the SiPM-based prototype modules, where SiPMs from three different manufacturers are used.\\

\end{abstract}


\end{frontmatter}


\section{Introduction}

Imaging Atmospheric Cherenkov Telescopes (IACT) capture faint images in Cherenkov light, lasting a few nanoseconds. These are produced in the atmosphere by very high-energy cosmic and gamma rays \citep{aharonian_astrophysics_2013}. Currently, the cameras of all relatively large-size IACTs are based on classical Photo Multiplier Tubes (PMTs) as light detectors \citep{mirzoyan_technological_2022,magic_collaboration_major_2016, aharonian_observations_2006, rajotte_upgrade_2014, mazin_large_2016}. SiPMs are being considered for the next generation of cameras for the Large-Sized Telescopes (LSTs) of the Cherenkov Telescope Array (CTA) \cite{heller_next_2023}. Smaller-size IACTs such as FACT or the prototypes of the Small-Sized telescopes (SSTs) and Schwarzschild–Couder telescopes (SCTs) of CTA use Silicon Photomultipliers (SiPMs) as light detectors \cite{mirzoyan_sipm_2013,anderhub_design_2013,biland_calibration_2014, benbow_status_2017, montaruli_small_2015, di_venere_prototype_2023}. The increasing popularity of SiPMs might suggest that also large-size IACTs could be based on SiPMs. However, there is no conclusive study that directly compares and measures the PMT and SiPM performances within large-size IACTs.
Instead, there is much speculation about the comparative performance of SiPMs and PMTs.

We developed three SiPM prototype detector modules at the  Max Planck Institute for Physics (Munich, Germany) \cite{hahn_development_2018}. These modules are adapted to the mechanical, electronic, and signal requirements to be installed in the imaging camera of the MAGIC-I telescope \cite{hahn_performance_2023}.\\
The imaging camera of a MAGIC telescope includes 1039 PMT-based channels (pixels), which are combined into clusters of seven \cite{magic_collaboration_major_2016}. The 1-inch size PMTs of enhanced quantum efficiency, hemispherical-shape input window, and 6-dynodes from Hamamatsu (type R-10408-01) are used in the cameras. The average peak quantum efficiencies of the bialkali photocathodes are 34 \% and 32 \% for MAGIC-I and MAGIC-II, respectively \cite{nakajima_new_2013}. The light-sensitive area of the camera has a round shape that fits the inner circle of the hexagonal shape cluster carrying mechanics. The six corners of the hexagon offer the possibility to install up to six additional detector modules, each of up to seven pixels. Each PMT is given a hexagonal-shaped light guide that collects light, minimizes the inter-pixel dead area, as well as rejects the stray light (albedo).\\
To be able to build a SiPM-based pixel that can replace the 1-inch PMT we used closely-packed matrices of seven or nine $6\times6\,\mathrm{mm}^2$ SiPMs. We actively sum up the signal outputs of the 7 (or 9) SiPM chips to mimic a composite sensor with an area equivalent to that of the PMT \cite{fink_sipm_2016, fink_second_2016}.\\
The first SiPM detector module, based on 7 pixels, each composed of 7 chips of $6\times6\,\mathrm{mm}^2$ size from Excelitas, was installed on May 9, 2015 in the MAGIC-I camera \cite{hahn_development_2017}. The experiences gained from its design and operation led to a second-generation design with improved sensitivity and stability. The latter modules were installed in 2017 \cite{hahn_results_2019}. All SiPM modules have been in regular operation since their installation in the MAGIC-I camera.\\
We regularly perform different calibrations and check the stability of the modules. Here we compare the SiPM and PMT performances directly, with no assumptions. For this, we use Cherenkov light flashes from cosmic-ray air showers. In addition, we flash nanosecond  (ns) short pulses from a laser operating at 355\,nm, installed in the center of the reflector.
The inclusion of realistic pixel designs, fill factors, electronics, and calibration procedures verified by various measurements allowed us to precisely assess the performances. Our results may show differences from other studies, probably due to their sometimes simplified assumptions see, for example, \cite{arcaro_study_2022}.

We describe the operational parameters of IACTs in Section 2, followed by an exploration of the general SiPMs pixel designs and light guides in Section 3. Section 4 elaborates on the design of the supply voltage and signal summation electronics. In Section 5, we calculate the expected responses and signal-to-noise ratios for both SiPM-based and PMT-based pixels, drawing from the spectra of the light of the night sky (LoNS) and airshower Cherenkov light. Additionally, in Section 5, we assess the linearity of our SiPM pixels and investigate their temperature stability. The installation of the SiPM modules into the MAGIC-I camera is briefly described in Section 6, while Section 7 elucidates various calibration procedures. Measured performances, utilizing artificial light pulses, airshower Cherenkov light from protons and muons, as well as analyses of time resolution and trigger performance, are discussed in Section 8. Section 9 covers the utilization of bandpass filters to mitigate the impact of LoNS. We summarize our findings in section 10.
\section{Operational conditions in IACTs}
\label{sec:operational_conditions}

In contrast to many particle physics experiments (see e.g., \cite{bornheim_integration_2023}), the cameras of IACTs operate at close to ambient temperatures. This has almost no influence on the performance of PMTs but SiPMs show a variable level of noise due to thermally excited dark counts \citep{renker_advances_2009, mirzoyan_sipm_2013}. We want to note that this contribution is significantly less compared to the photo-electron rate induced by LoNS.\\
For installation of a SiPM prototype module in the MAGIC-I imaging camera housing one needs to cope with a number of constraints such as the size, required power, heat dissipation, and electronic noise. 
Because the project's aim is to directly compare SiPM to PMT-based pixels, we use the same light guide entrance aperture and the number of pixels per module.
\\
All detector modules are connected to two cooling plates which are centered in the camera housing \citep{nakajima_new_2013}. The dissipated heat of the light sensors and electronics is conducted via the module's metal structure to those cooling plates. These cooling plates use a closed circuit of liquid coolant to transport the excess heat from the modules to the outside of the camera. Typically we stabilize the camera temperature not too far from the ambient temperature. In the SiPM prototype modules, we use several sensors to monitor the temperature at different locations across the module. Our study of operation temperature and stability is described in section \ref{sec:thermal}.\\
The Domino Ring Sampler version 4 (DRS4) \citep{bitossi_ultra-fast_2016} based readout of the MAGIC telescopes is located in the counting house of the experiment. The analog electronic signals of the light sensors are converted to analog optical signals by fiber-coupled vertical-cavity surface-emitting laser (VCSEL) diodes and then transmitted to the readout via $162\,\mathrm{m}$ long multimode optical fibers \citep{magic_collaboration_major_2016,michalzik_operating_2003}. The sampling frequency of the DRS4 readout is set to $1.64\,\mathrm{GHz}$ and stored events consist of 50 samples, resulting in a waveform of about $30\,\mathrm{ns}$ per event. The SiPM prototype modules provide analog optical output for each pixel and are connected to spare channels of the MAGIC readout system.
\section{Sensors and light guides}
\label{sec:sensors}

\subsection{Types of used SiPMs}
\begin{figure*}[htb]
\begin{subfigure}[b]{0.3\textwidth}
    \centering
    \includegraphics[width=1.08\textwidth]{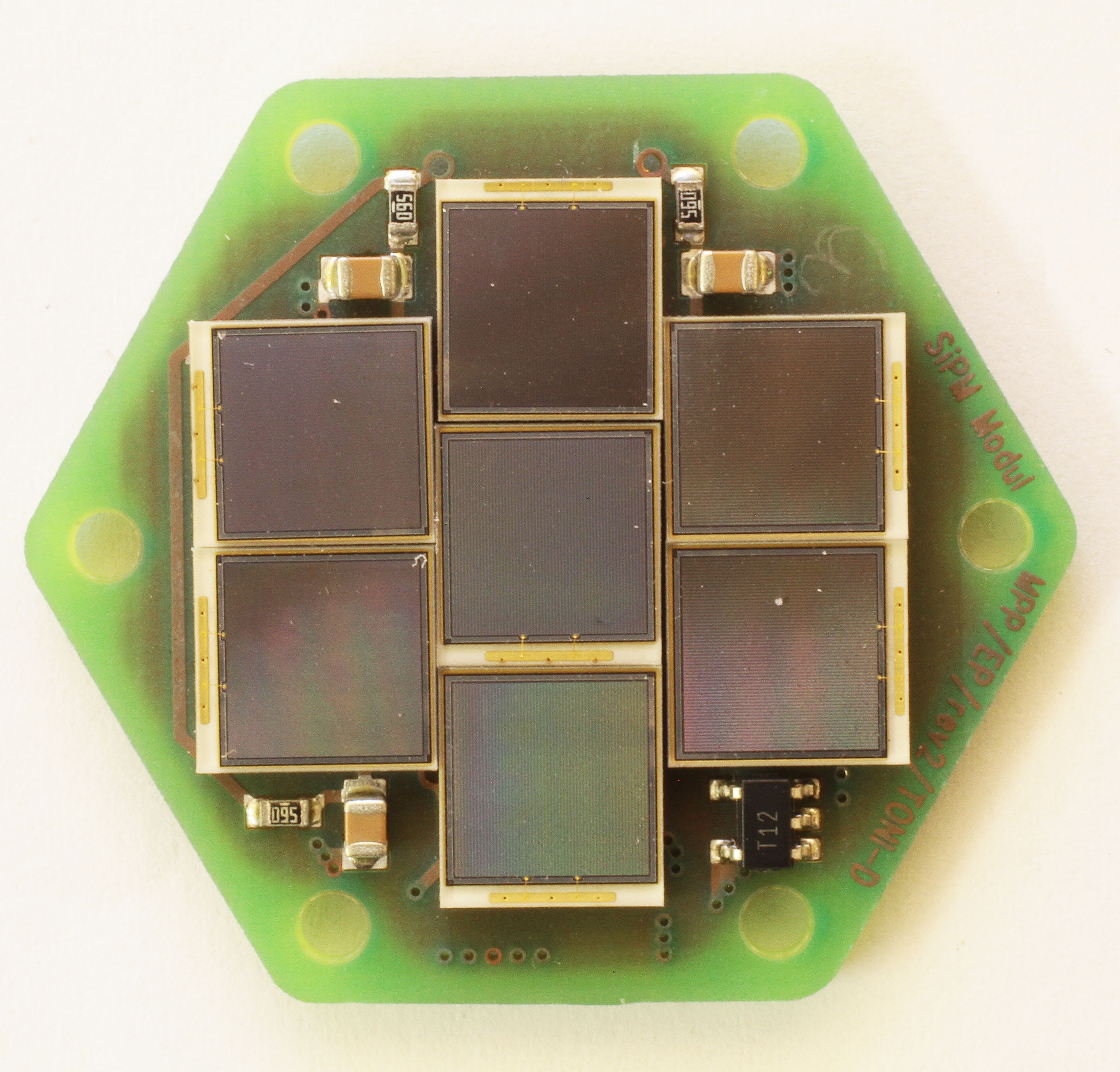}
    \caption{}
    \label{fig:pixel_excelitas}  
\end{subfigure}
\hspace{2.5mm}
\begin{subfigure}[b]{0.3\textwidth}  \centering  
    \includegraphics[width=1.08\textwidth]{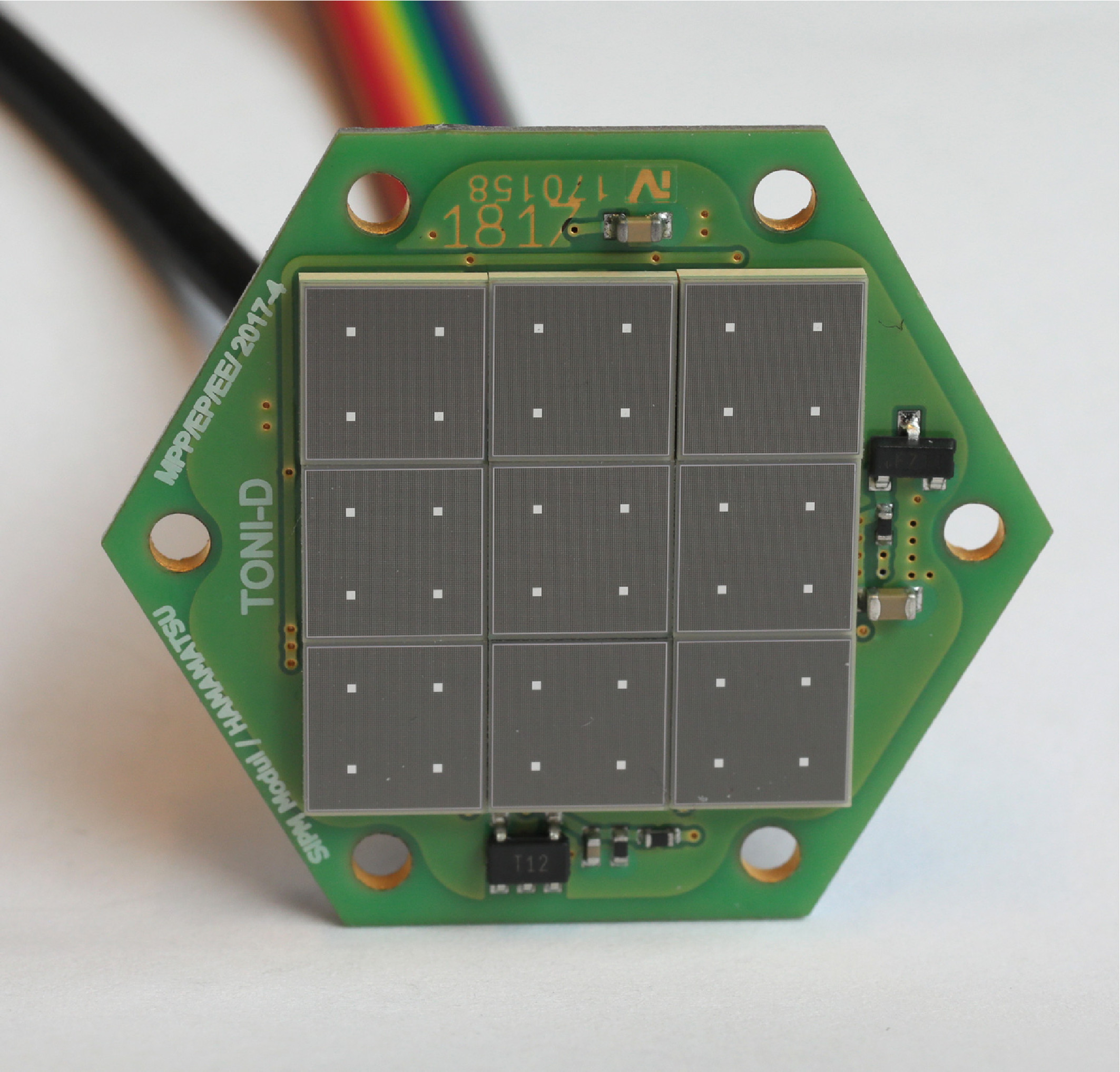}
    \caption{}
    \label{fig:pixel_hamamatsu} 
\end{subfigure}
\hspace{2.5mm}
\begin{subfigure}[b]{0.3\textwidth}
    \centering
    \includegraphics[width=1.08\textwidth]{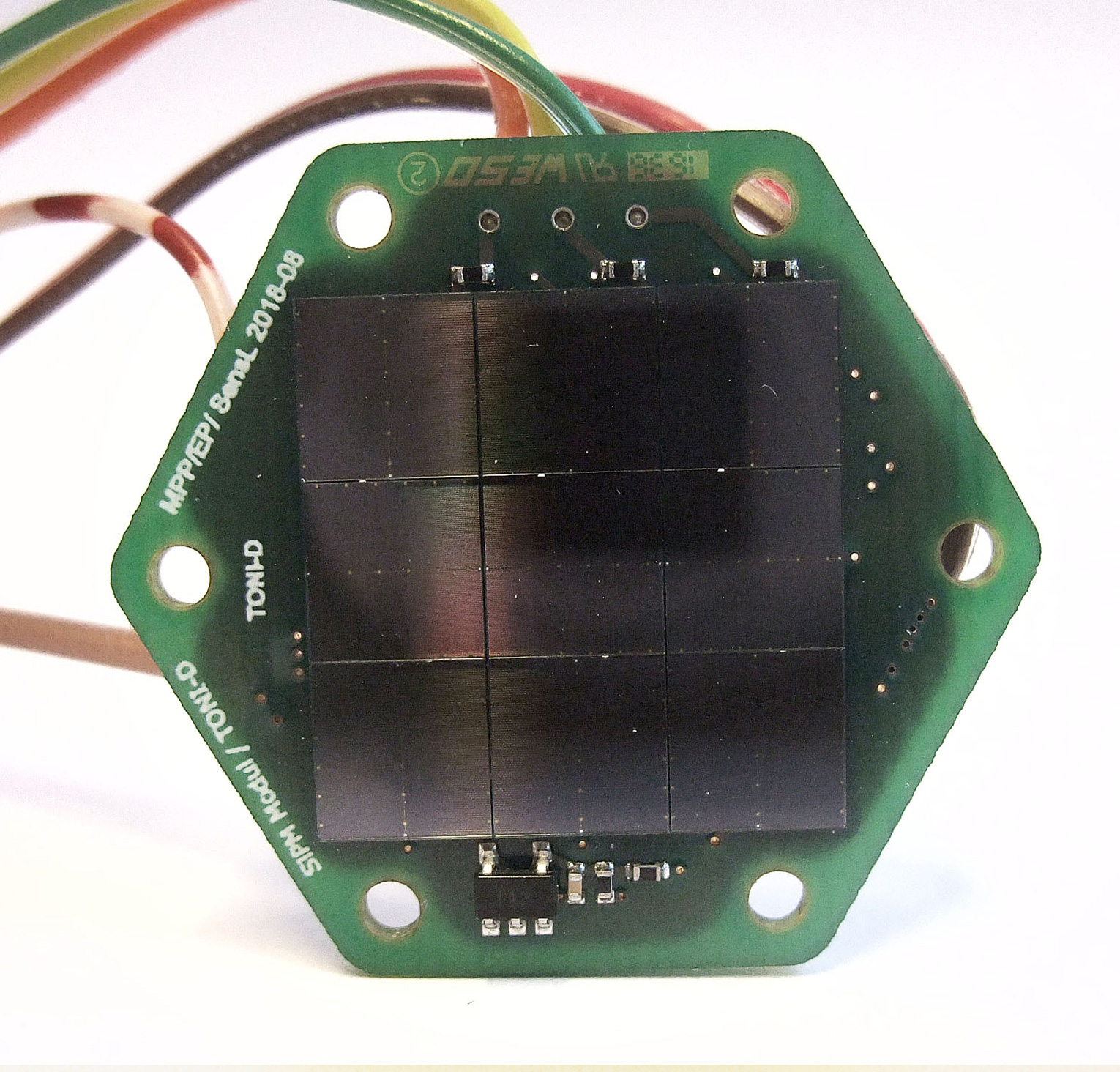}
    \caption{}
    \label{fig:pixel_SensL} 
\end{subfigure}
\caption{The first generation SiPM pixel is based on Excelitas SiPMs (a). Two second-generation pixels, one based on Hamamatsu (b) and the other one on SensL SiPMs (c). Images from \cite{hahn_results_2019}.
}
\label{fig:pixels}
\end{figure*}

For the first prototype, we chose Excelitas SiPMs of type C30742-66-050-X of size $6\times6\,\mathrm{mm^2}$ and $50\,\mathrm{\mu m}$ cell pitch. Each pixel of this first-generation module uses seven SiPMs. The second generation modules followed two years after the first module. These were based on S13360-6075VS Hamamatsu SiPMs ($6\times6\,\mathrm{mm^2}$, $75\,\mathrm{\mu m}$ cell pitch) and MicroFJ-60035-TSV SiPMs from SensL ($6\times6\,\mathrm{mm^2}$, $35\,\mathrm{\mu m}$ cell pitch). The second-generation pixels use a matrix of nine SiPMs. All three pixel designs are shown in figure \ref{fig:pixels}.

\subsection{Acceptance and angular cut-off}
\label{sec:acceptance}

\begin{figure}[tb]
\centering
\includegraphics[width=1\columnwidth]{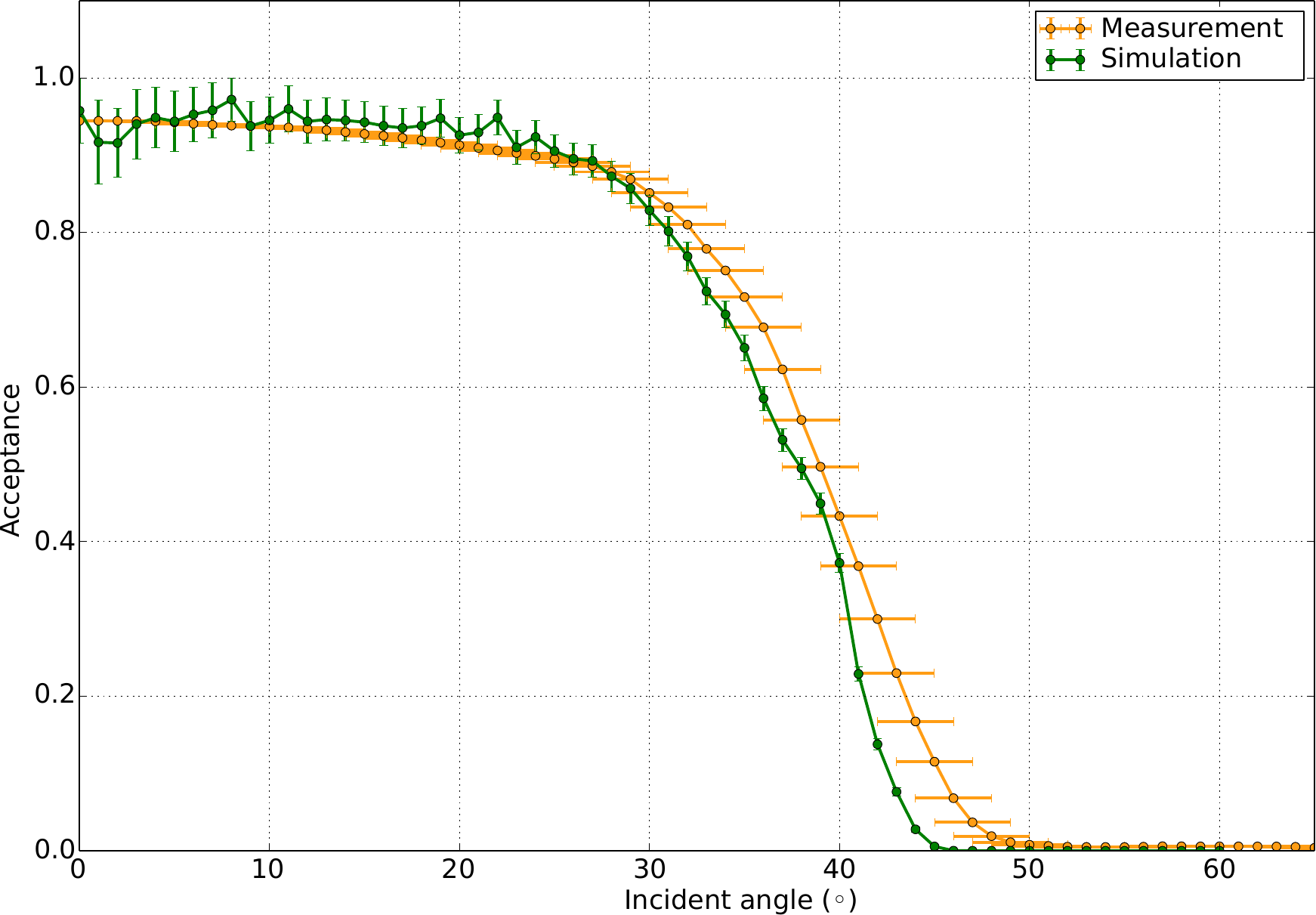}
\caption{Comparison of simulated and measured acceptance of the light guide. Uncertainty bars include systematic and statistical errors.
}
\label{fig:Light_guide}
\end{figure}

The pixels of an IACT are coupled to light guides. These have a twofold purpose: a) they minimize the dead area between the pixels by concentrating the light to a small active area, and b) albedo reduction ---  reject incident light arriving from beyond the telescope's mirror dish. Even though we targeted a pixel with an area similar to that of a PMT, we could not use the same light guides because these were optimized for a PMT with a hemispherical photocathode shape but the SiPMs are flat. In addition, SiPMs are only sensitive to light arriving at an incident angle of less than $70^{\circ}$ due to Fresnel reflection \cite{saleh_fundamentals_2007}. For the given topology of the MAGIC telescope, we calculated an optimal cut-off angle of $31.22^{\circ}$ from the normal to reject ambient light. More details on the incident angles and the optimized light guide can be found in \cite{hahn_development_2017}.\\
Under these boundary conditions, the light concentrator's maximum useful compression ratio is $3.28$ for a hollow light guide. Each side of our hexagonal light guides is parabolic and covered with a high reflectivity aluminized mylar film. We optimized the acceptance (i.e.~ratio of incident photons to photons reaching the detector surface) of our light guides with simulations using the ray-tracing library ROBAST \cite{okumura_development_2011, okumura_robast:_2015}. This optimization took the spectrum of the air shower Cherenkov light and the wavelength-dependent reflectivity of the aluminized mylar into account.\\
The simulations yield a collection efficiency of $91.8\,\%\pm0.3\,\%$. We measured the collection efficiency to be $92.1\,\%\,\substack{+\;0.7\,\% \\ -\;1.0\,\%}$ for light that can be detected by the SiPM. This efficiency of our adapted light concentrator design is within 1\,\% of the standard MAGIC light concentrator efficiency when mounted on the hemispherical PMT cathode for which it was optimized.
Simulation and measurement of the acceptance of the produced light guide are shown in figure \ref{fig:Light_guide}. They show agreement within uncertainties with an absolute difference of only $0.3\,\%$. The discrepancy between simulation and measurement at around $45\degr$ is likely caused by manufacturing imperfections and potential dust deposits on the reflective surface which play a more dominant role at the shallow angles under which reflections occur at these incident angles.

\subsection{Pixel dead area}
\label{subsec:pixel_dead_area}

\begin{figure}[tb]
\centering
\includegraphics[width=0.99\columnwidth]{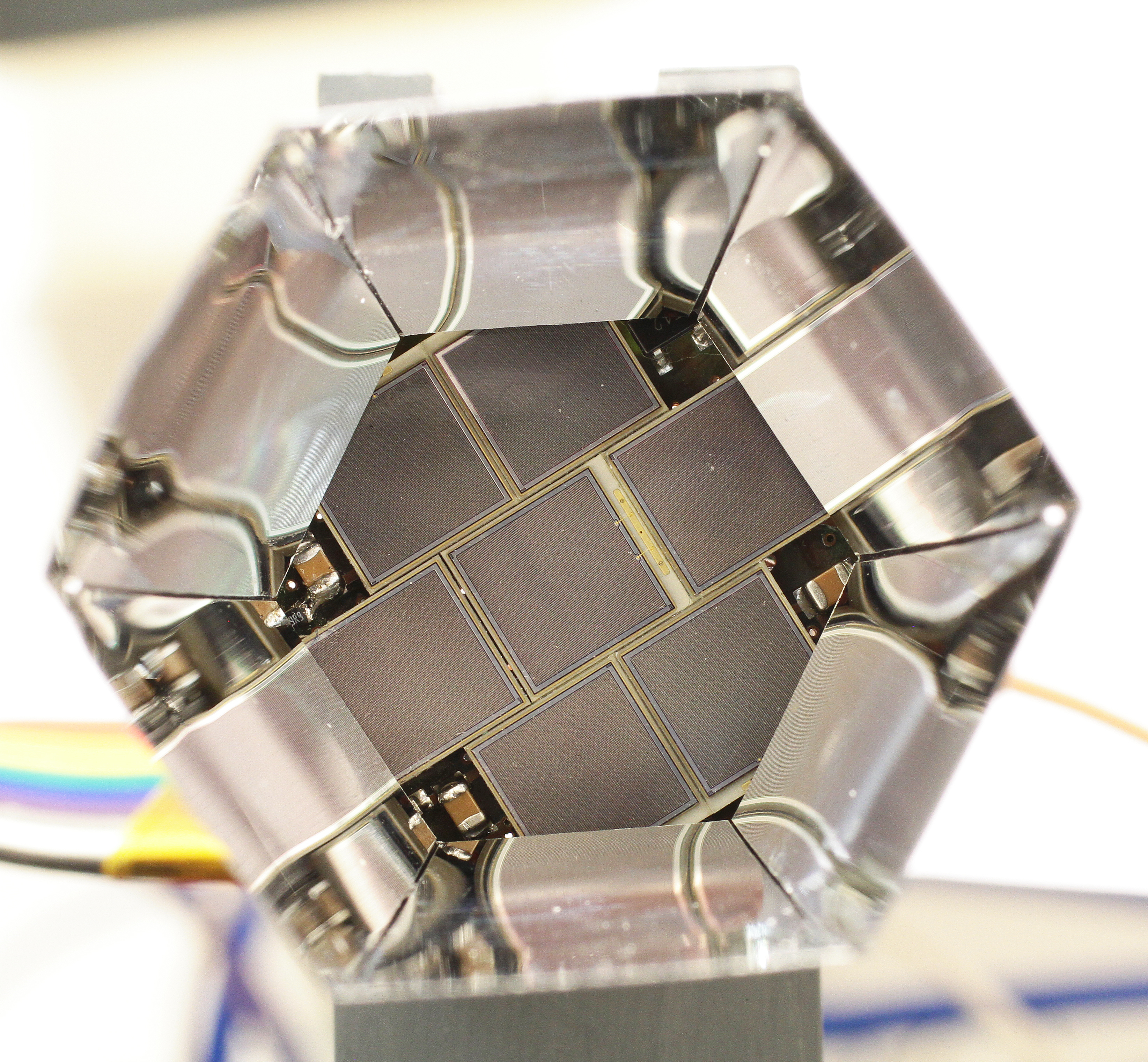}
\caption{First generation pixel (Excelitas) with mounted light guide, top view. The dead area of the SiPM packaging, between the individual SiPMs and at the corners can be clearly seen.
}
\label{fig:Excelitas_pixel}
\end{figure}

\begin{figure}[htb]
\centering
\includegraphics[width=0.99\columnwidth]{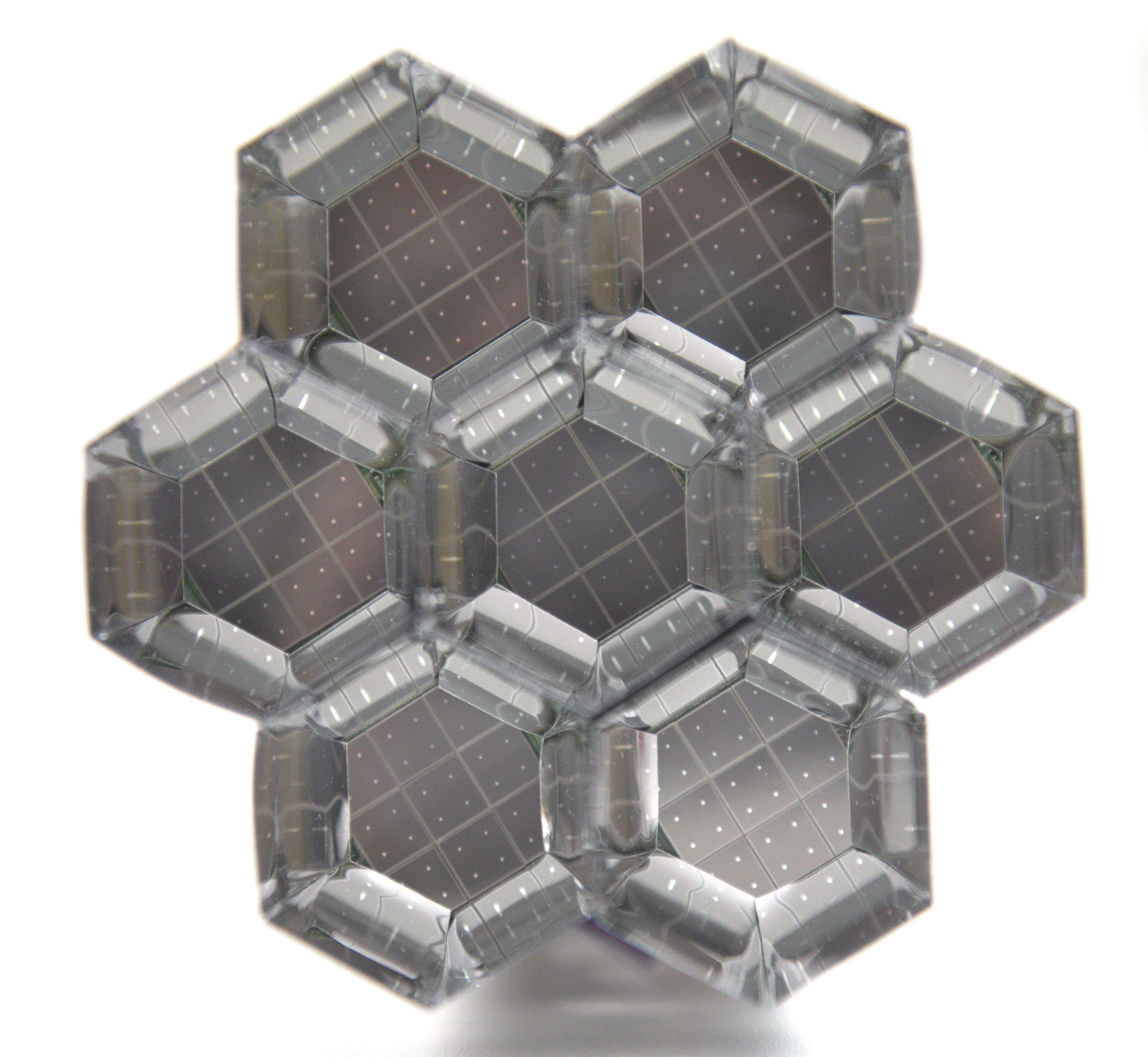}
\caption{Hamamatsu SiPM prototype module with seven SiPM pixels with light guides, top view.}
\label{fig:Hamamatsu_module}
\end{figure}

The SiPMs of one pixel are soldered onto a printed circuit board (PCB). Dead area of the pixel is caused by the packaging of the SiPM as well as small gaps that are left between the sensors after the reflow soldering process. In addition, the first generation of SiPM pixels using Excelitas SiPMs had four dead areas in the corners where electronic components were soldered. This can be clearly seen in figures \ref{fig:pixel_excelitas} and \ref{fig:Excelitas_pixel}. By increasing the number of SiPMs to nine for the second generation of modules, these dead corners were eliminated, see figures \ref{fig:pixel_hamamatsu} and \ref{fig:pixel_SensL}. We show a complete module of seven second-generation pixels in figure \ref{fig:Hamamatsu_module}.

Only part of the active area of the SiPMs is exposed to light through the light guide. This has no significant implications as the inaccessible active area contributes only little to the total count rate (see further discussion in section \ref{sec:expectations_spectra}).

Using microscope images we measured the fraction of active area taking into account the above-mentioned effects to be $69\,\%$ for the Excelitas pixels which was acceptable for a first proof of principle, $87\,\%$ for the SensL and $84\,\%$ for the Hamamatsu SiPM pixels.
This does not include the fill factor of the SiPMs, since it is already included in the PDE of the devices. These measured fractions are also listed in table \ref{tab:pixel_cell_recharge}.

\section{Electronics}
\subsection{HV and offset voltages}

The Excelitas SiPMs of the first module showed a wide spread in their breakdown voltages ranging of about $13\,\mathrm{V}$. To be able to set the bias voltage to all the seven SiPMs of a pixel in an economic way we combined the individual SiPMs into three groups, containing 2, 2, and 3 sensors each. The SiPMs for a given group were selected to have similar breakdown voltages.\\
During the later construction of the Hamamatsu and SensL SiPM pixels, we kept this combination into three groups of consequently three SiPMs each, although the low spread in the breakdown voltages, between $0.04\,\mathrm{V}$ and $0.14\,\mathrm{V}$ per pixel, would not have made this necessary. The reason is that this subdivision of a pixel, apart from setting equal bias voltages to each group, enabled us to switch parts of a pixel on and off.

\begin{figure}[htb]
\centering
\includegraphics[width=1\columnwidth]{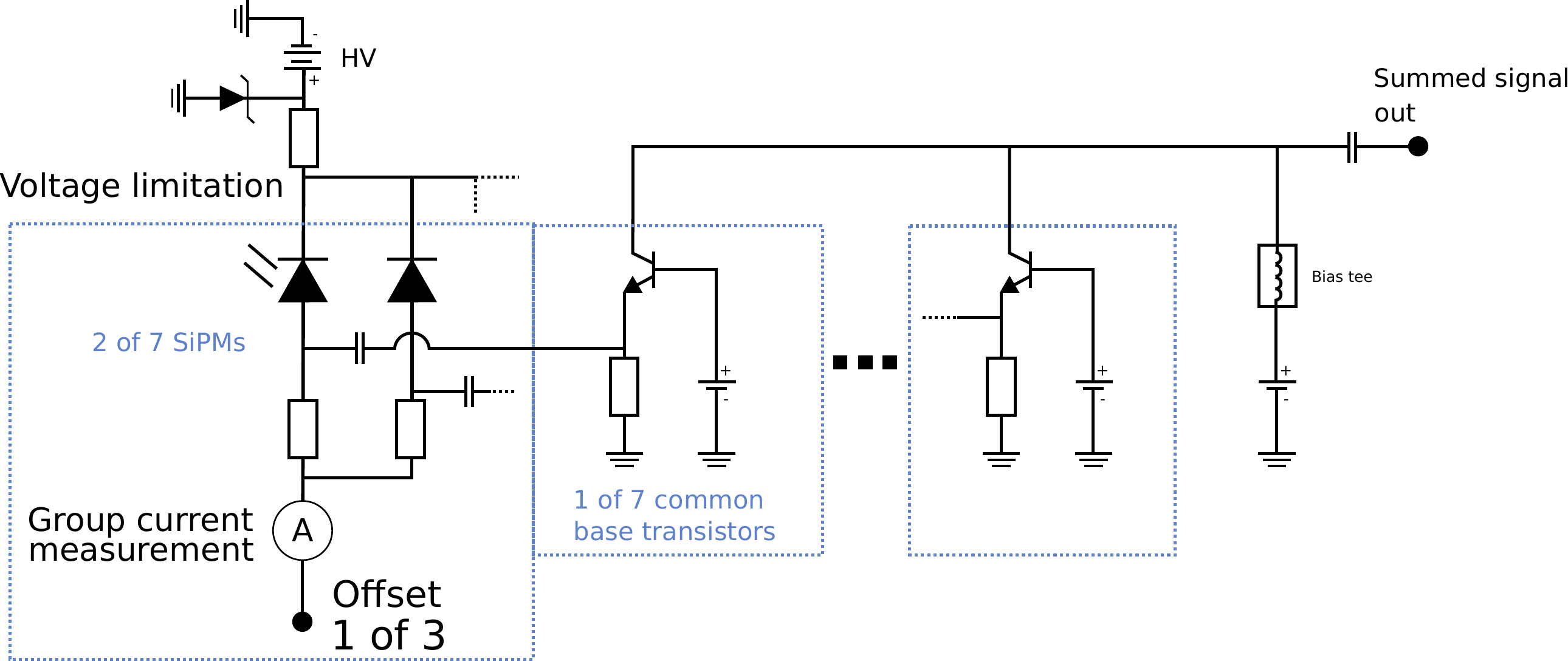}
\caption{Schematic circuit diagram.}
\label{fig:schematics}
\end{figure}

All pixels of a module share a single relative high voltage (hereafter HV) power supply. The bias voltage of each SiPM group of each pixel is adjusted by applying an offset voltage on the anode side of each SiPM group. 
The total bias voltage is then the difference between the HV and offset voltage.
The simplified schematic circuit can be seen in figure \ref{fig:schematics}.\\
If the current limiting resistor in the HV path is chosen too high, it would lead to a varying bias voltage and, consequently, SiPM gain with changing SiPM currents, rendering any calibration unreliable. To circumvent this we chose a very small resistor of only $56\,\Omega$ ($120\,\Omega$ in the case of Excelitas).
This way the HV variations stay below $10\,\mathrm{mV}$ for all realistic illumination scenarios. The original idea was to use an inductor, but noise pickup proved to be better attenuated by the combination of the small series resistance and the device's capacitance.

\subsection{Summation} 
\label{sec:electronics_summation}

Currently, the size of a genuine (not composite) SiPM is limited to not more than 6-7 mm edge length. For a SiPM of larger size, one can use several SiPMs and sum up their outputs. A simple parallel connection of SiPMs will suffer from the large capacitance and consequently low bandwidth, and slow response time. 
To keep the original fast response time of the used single SiPM chips, the outputs of the SiPMs shall be “isolated” from each other before summation. This can be done, for example, by using a single transistor decoupling "common base" stage with low input and high output impedances as shown in figure \ref{fig:schematics}.
A detailed description of the circuit design can be found in \cite{fink_sipm_2016} and \cite{fink_second_2016}.

\begin{figure}[h]
\centering
\includegraphics[width=1\columnwidth]{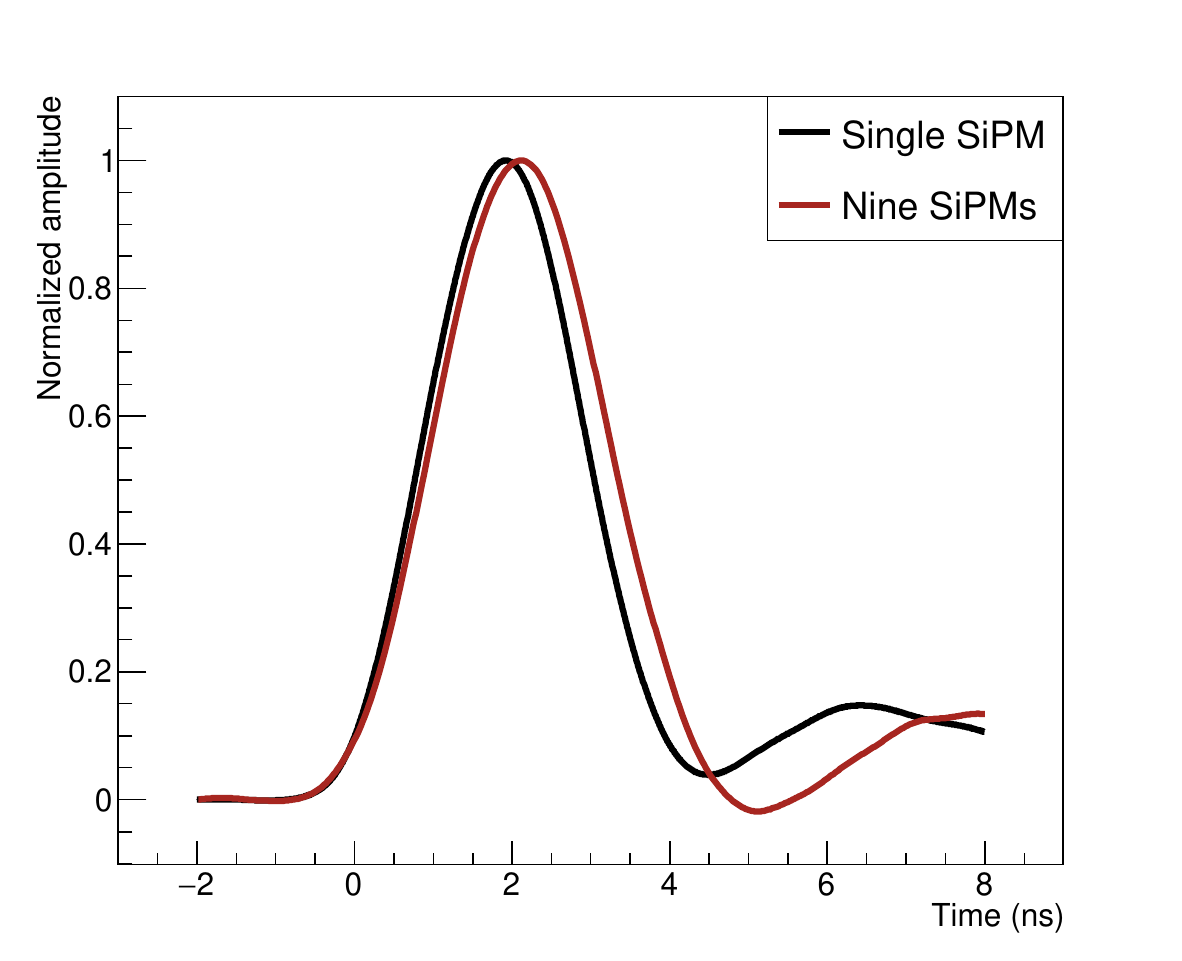}
\caption{Comparison of the pulse shape of a single SiPM chip with the active sum of nine Hamamatsu SiPM chips. The pulse FWHM widens only slightly from $2.4\,\mathrm{ns}$ to $2.8\,\mathrm{ns}$. The amplitudes were normalized for better visualization.}
\label{fig:waveform_comparison}
\end{figure}

The second generation of SiPM-based pixels uses a similar transistor circuit design. In figure \ref{fig:waveform_comparison} we compare the average pulse shapes of one single Hamamatsu SiPM with the summed output signal of the nine SiPMs of one pixel. The pulses shown are the amplitude normalized averages of 85,000 light-emitting diode (LED) flasher events sampled at a resolution of 10\,ps for the single SiPM and 174,000 events sampled at a resolution of 20\,ps for nine SiPMs. This results in negligible systematic and statistical uncertainties for this measurement.
The output pulse full width at half maximum (FWHM) increases only slightly from $2.4\,\mathrm{ns}$ to $2.8\,\mathrm{ns}$ for Hamamatsu SiPM-based pixels and similarly from $3.1\,\mathrm{ns}$ to $3.4\,\mathrm{ns}$ for pixels based on SensL SiPMs. For the first generation SiPM-based pixels using Excelitas SiPMs, the FWHM increased from $5.1\,\mathrm{ns}$ to $5.9\,\mathrm{ns}$ \citep{hahn_development_2017}.

\section{Expected LoNS and Cherenkov light responses}

\subsection{Spectra and integrals}
\label{sec:expectations_spectra}

\begin{figure}[]
\begin{subfigure}[htb]{0.84\columnwidth}
    \centering
    \includegraphics[width=1.1\textwidth]{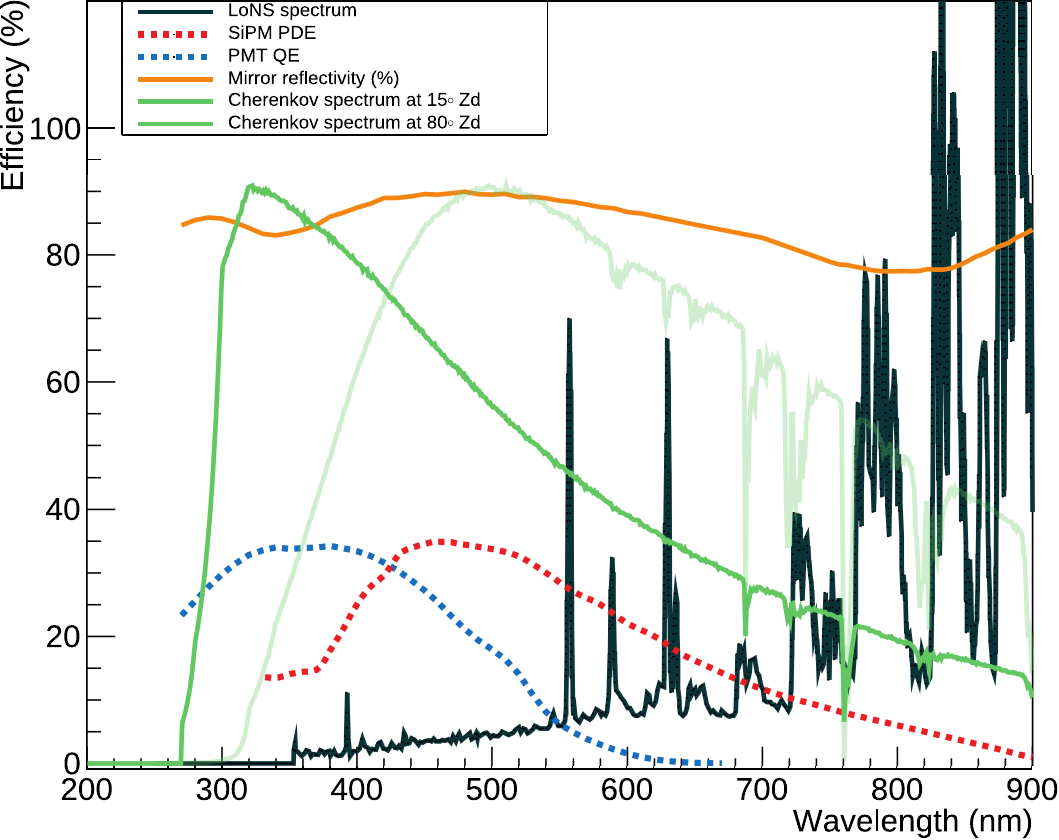}
    \caption{}
    \label{fig:spectra_excelitas}  
\end{subfigure}\\
\begin{subfigure}[htb]{0.84\columnwidth}
  \centering  
    \includegraphics[width=1.1\textwidth]{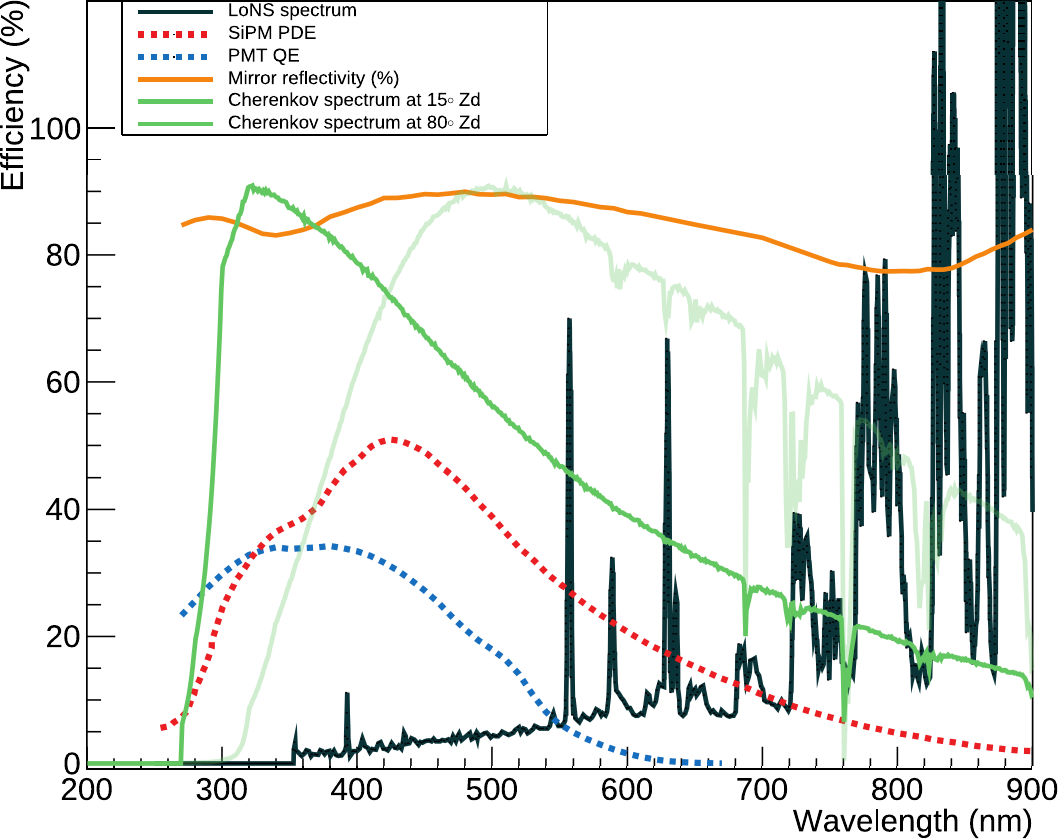}
    \caption{}
    \label{fig:spectra_hamamatsu} 
\end{subfigure}\\
\begin{subfigure}[htb]{0.84\columnwidth}
    \centering
    \includegraphics[width=1.1\textwidth]{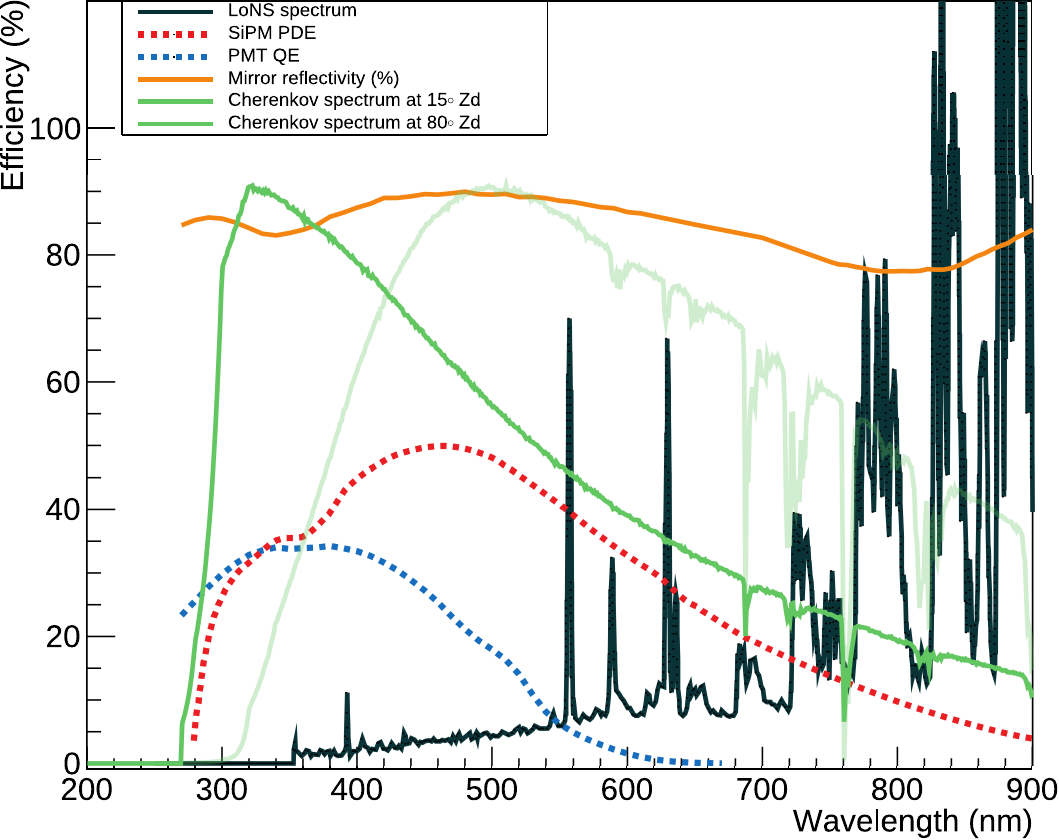}
    \caption{}
    \label{fig:spectra_SensL} 
\end{subfigure}
\caption{PDEs (Excelitas (a), SensL (b), Hamamatsu (c)) shown in red, QE of MAGIC PMTs shown in blue, LoNS spectrum (black, in arbitrary units (A.U.))
and Cherenkov spectra of air showers observed under 15\degr (green, in A.U.) and 80\degr (light green, in A.U.) zenith distance. The zenith dependency of the LoNS can be seen in figure \ref{fig:DC_vs_Zd}. The Cherenkov spectra are generated from Monte-Carlo simulations summing all azimuth angles. The mirror reflectivity of the MAGIC-1 mirrors is shown in orange. Cherenkov and LoNS spectra were scaled for better visibility. 
}
\label{fig:PDE_LoNS_Cher}
\end{figure}

To estimate the expected performance of our SiPM prototype modules we compared the spectral photon detection efficiency $\mathrm{PDE}(\lambda)$ curves of light sensors with the spectra of the LoNS and the Cherenkov light from air showers reaching the telescope camera. The spectra are shown in figure \ref{fig:PDE_LoNS_Cher}. In recent years, MAGIC observations' mean zenith distance (ZD), which denotes the angular distance between the telescope pointing direction and zenith, has steadily increased to cover a larger solid angle with more astrophysical sources and a wider energy range.
For an example see \cite{acciari_magic_2020}. Therefore, it is important to compare the detector's PDE with the Cherenkov and LoNS spectra not only close to the zenith but also at large zenith distances.\\ 
The Cherenkov light spectrum of air showers at the telescope camera peaks at around $320\,\mathrm{nm}$ for low zenith observations, at an observational altitude of 2.2 km. The peak quantum efficiency (QE) of PMTs makes a good match with the Cherenkov light spectrum. All of the SiPMs we used have their peak sensitivity at longer wavelengths, thus detecting a lower number of Cherenkov photons from the main peak in the near UV region. On the other hand, these collect more photons from the long-wavelength tail of the Cherenkov spectrum. At large and very-large zenith distances the peak of the Cherenkov light spectrum shifts towards the blue-green region. This is due to the longer distance to the shower maximum region, from where the majority of photons arrive at the telescope. The longer paths attenuate the near-UV part of the spectrum, due to the Rayleigh scattering and absorption effects. 
The Cherenkov spectra were generated with CORSIKA \cite{heck_corsika_1998} and the photon propagation to the IACT camera was modeled with an adapted version of MARS, the MAGIC Analysis and Reconstruction Software \cite{moralejo_mars_2009}.
We checked the dependency of the Cherenkov spectra on the impact parameter and energy for different airmasses. The impact parameter refers to the distance between the telescope and the point where the shower axis intersects the ground, while airmass quantifies the ratio of atmospheric column density along the line of sight compared to the column density when observing at zenith. We found that the Cherenkov spectrum has slightly more UV light for showers with an impact parameter of 0\,m at low zenith distances and for low and medium zenith distance showers with energies $\gtrsim15$\,TeV. This is of course to be expected because the Cherenkov photons have to traverse less atmosphere which leads to less absorption of the UV light. A semi-empirical model of the electromagnetic shower development as a function of atmospheric depth can be found in \cite{k_greisen_progress_1956}.
In reality, most triggered events have an impact parameter of 100 -- 150\,m, and hardly any events have an impact of 0\,m for low and medium zenith distances, even before the stereo reconstruction and the quality selection. The count rates of astrophysical sources at energies $>10$\,TeV are universally expected to be very low, making it more feasible to observe such high energies at very large zenith distances, where the collection area is about one order of magnitude larger than at low zenith distances \cite{mirzoyan_extending_2020, acciari_magic_2020}. For large and very large zenith distance air showers (airmass $\gtrsim$ 3 (AM3)) there is no dependency of the spectral shape on the energy or impact parameter.
We, therefore, simulate air showers of 5\,TeV gamma rays to achieve a high photon statistic per simulated event, with impact parameters between 0 -- 600\,m (same as used in the standard MAGIC Monte-Carlo (MC) simulations), randomly distributed along $2\pi$ in azimuth. We note that the increase in the simulated impact parameter range is not relevant for the determination of the Cherenkov spectra, but only for a potential calculation of the effective collection area of the telescope. This is very similar to the approach used in \cite{arcaro_study_2022}.
The shown spectra in figure \ref{fig:PDE_LoNS_Cher} are the result of several thousand of such simulated events at zenith distances 10\degr and 80\degr.
We considered all photons reaching the imaging camera and did not include any simulation of a trigger logic, since this happens only after the photons are converted into electronic signals.

\begin{figure}[tb]
    \centering
    \includegraphics[width=1\columnwidth]{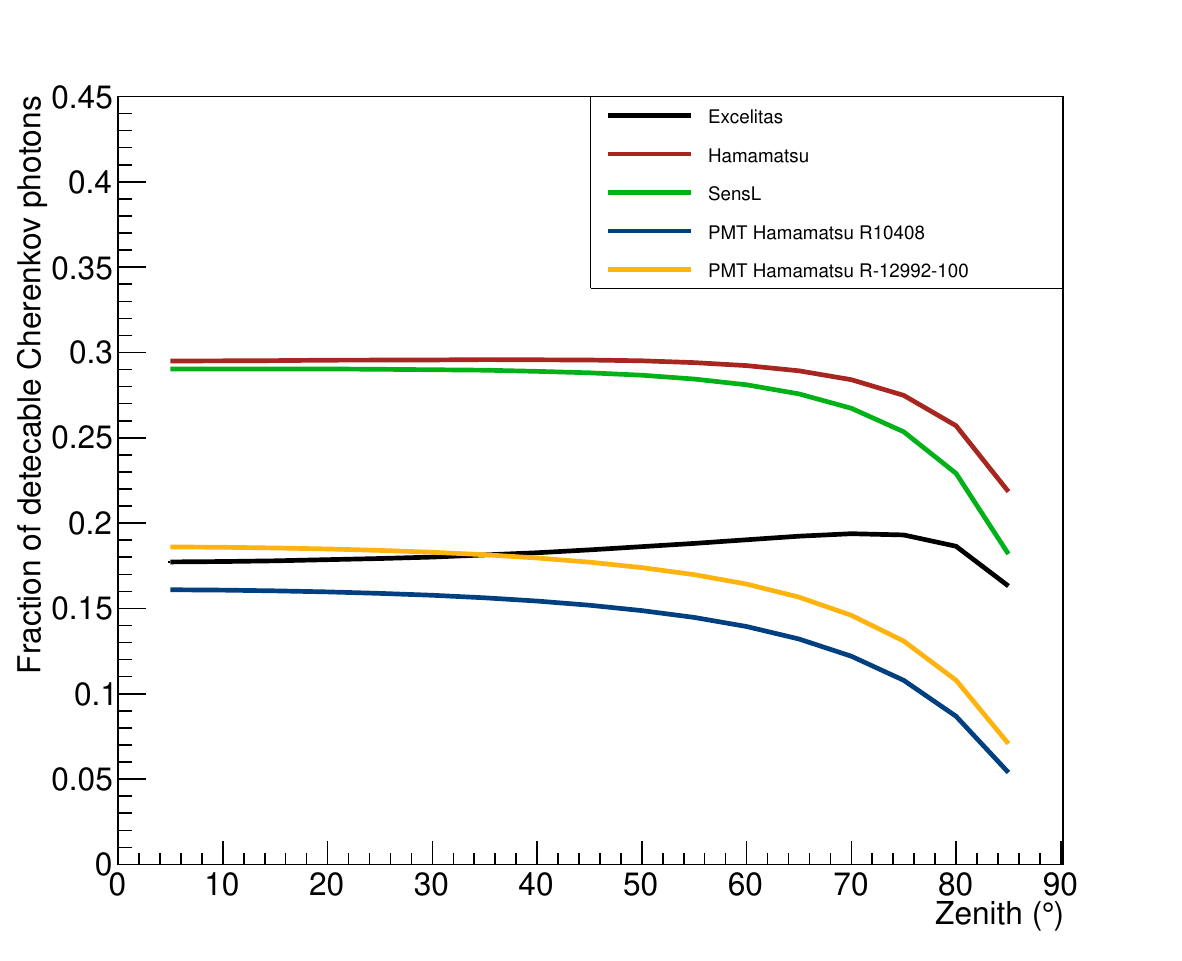}
    \caption{Fraction of the detectable Cherenkov photons in the range of 250\,nm to 900\,nm with a given photosensor after atmospheric absorption, mirror reflectivity, and camera entrance window. Effects like the pixel fill factor or possible total reflection at the sensor's surface for photons with low incident angles are not included here. }
    \label{fig:Signal_vs_Zd}
\end{figure}

By multiplying the detection efficiency of a given detector with the wavelength-dependent mirror reflectivity, the absorption of the camera entrance window, and the Cherenkov spectrum of a given zenith distance one can calculate the detectable fraction of Cherenkov photons for this detector type. This represents the maximum achievable efficiency as it does not take into account camera and pixel fill factors. This fraction is shown in figure \ref{fig:Signal_vs_Zd} for the four different light detectors under study. In addition, we also show the resulting graph for the recent 7-dynode Hamamatsu PMT R12992-100-05 as measured by \cite{mirzoyan_evaluation_2017} and used for the medium and large-size telescope imaging cameras of CTA.
It is interesting to note that the SiPMs show a more flat behavior compared to the PMTs. All sensors show an efficiency drop once the majority of the Cherenkov light distribution is shifted towards longer wavelengths than the optimal PDE range.

\begin{figure}[tb]
    \centering
    \includegraphics[width=1\columnwidth]{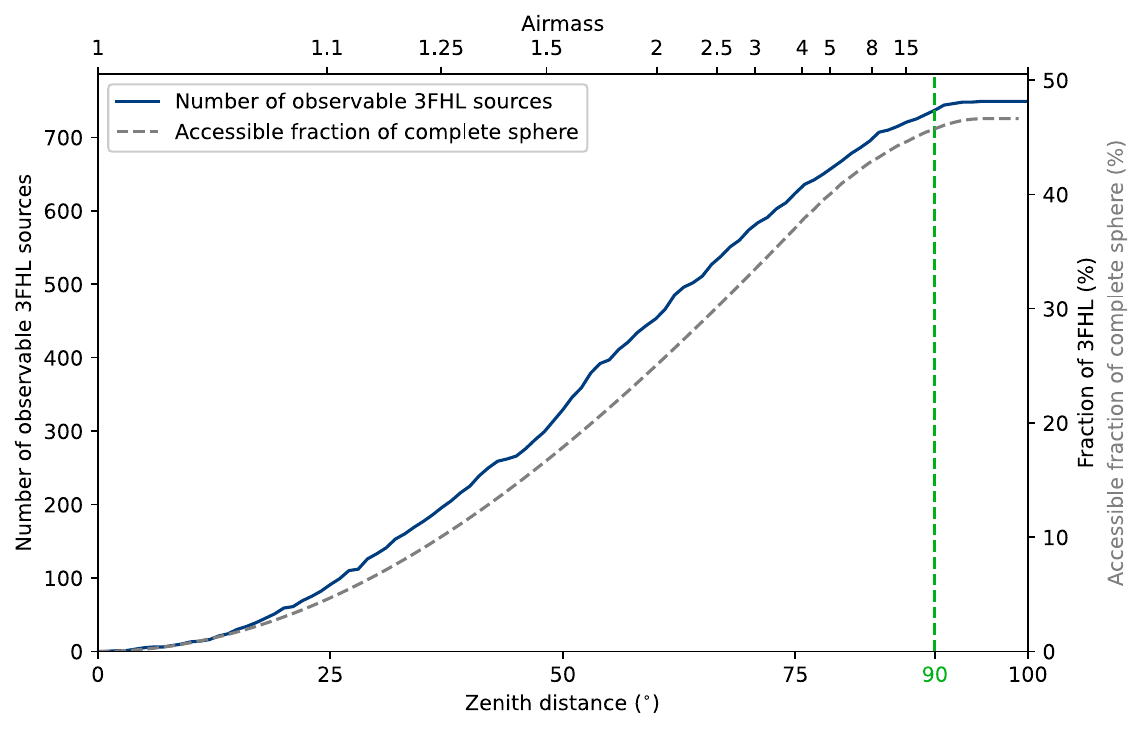}
    \caption{Number of 3FHL sources observable by MAGIC for all possible zenith distances (blue). Towards the North, the orography of the site allows one to observe at zenith distances larger than 90\arcdeg \cite{ahnen_limits_2018}. For reference, the accessible fraction of a complete sphere at the given zenith distance taking into account the orography is shown as a grey dashed line.
    }
    \label{fig:3FHL_sources}
\end{figure}

The main reason for building a large IACT is to achieve a low energy threshold. Especially low or medium zenith distance observations can provide a low threshold. On the other hand, over the course of recent years MAGIC has maximized observations at large zenith distances, see for example \cite{acciari_teraelectronvolt_2019, acciari_magic_2020, mirzoyan_extending_2020}. These observations can allow one to access sources in a wider solid angle as well as to extend the energy range of MAGIC towards higher energies, from multiple tens of TeV to above 100 TeV. To further motivate the necessity of larger zenith distance observations we compute the number of observable point sources from the Third Catalog of Hard \textit{Fermi}-LAT Sources (3FHL) \cite{ajello_3fhl_2017} from the MAGIC telescopes' site. This evaluation gives an even better estimate than just calculating the observable fraction of the celestial sphere because 
it takes into account the distribution of high-energy gamma-ray sources, which leads to a steeper rise in mid-zenith range.
For the sake of simplicity, we ignore the potential existence of a spectral cut-off.
By increasing the maximum zenith distance for observations from 50\arcdeg to 85\arcdeg, the fraction of observable 3FHL sources more than doubles from 21\% to 46\%. Similar numbers can be obtained using the more recent fourth \textit{Fermi} Large Area Telescope (LAT) catalog of $\gamma$-ray sources (4FGL-DR3) \cite{abdollahi_incremental_2022}. Here we ignored possibly curved spectra of the sources, the reason as to why many sources are not detectable or only detectable below a given zenith distance in the IACT energy range. Further assumptions on the flux levels and spectral shapes, necessary for accurate calculations, are beyond the scope of this paper. The number of observable 3FHL sources is shown in figure \ref{fig:3FHL_sources} for all possible zenith distances at which MAGIC can observe while taking into account the surrounding landscape.

 After taking into account the pixel fill factors in the imaging camera, the spectral responses can be integrated and allow a direct comparison between different sensors. For low zenith distance observations this yields that with the first generation pixels based on the 10-year-old Excelitas SiPMs one will detect about $25\,\%$ less Cherenkov photons than with the MAGIC PMTs.
On the contrary, with the Hamamatsu and SensL SiPMs, the telescope will detect about $70\,\%$ and $60\,\%$ more Cherenkov photons than with the MAGIC PMTs used.

The emission spectrum of LoNS increases towards longer wavelengths and shows strong emission lines. For example, one can clearly see the atomic Oxygen line $\mathrm{O}({}^{1}\mathrm{S}-{}^{1}\mathrm{D})$ at $557.7\,\mathrm{nm}$ above a rising pseudo-continuum starting from the atmospheric cut-off at about $300\,\mathrm{nm}$. For more details on the LoNS spectrum, we refer the reader to \cite{roach_light_1973, patat_dancing_2008, barentine_night_2022}. The LoNS spectrum at the MAGIC site was measured in the range from $350\,\textrm{--}\,950\,\mathrm{nm}$ by \cite{benn_palma_1998}. For the ranges $\lambda\le350\,\mathrm{nm}$ and $\lambda\ge950\mathrm{nm}$ we used the \textit{SKYCALC Sky Model Calculator}\footnote{\burl{https://www.eso.org/observing/etc/bin/gen/form?INS.MODE=swspectr+INS.NAME=SKYCALC}} using the Cerro Paranal Sky Model for a moonless night at the La Silla site \cite{noll_atmospheric_2012, sanchez_night_2007}. 
Since the LoNS spectrum by \cite{benn_palma_1998} was measured at a low zenith distance we used the relative changes of each wavelength with zenith distance from \textit{SKYCALC} to extrapolate the measured La Palma LoNS spectrum to higher zenith distances. The uncertainty of this extrapolation is wavelength dependent and increases especially for wavelength $<500\,\mathrm{nm}$ and zenith distances larger than 60\degr \citep{noll_atmospheric_2012}. In the future, we plan to perform dedicated LoNS spectrum measurements at La Palma for different zenith distances.
The extrapolation is done in airmass space instead of zenith distance space using the formula for determining the airmass at a given zenith distance from \cite{keith_a_pickering_southern_2002}, although we find that the improvements compared to extrapolation in zenith distance space are marginal.
Here it is worth mentioning that the quantitative change of the LoNS background rate at 532\,nm was measured with the MAGIC atmospheric LIDAR system in \cite{fruck_characterizing_2022}. Given the respective resolutions, our model agrees with \cite{fruck_characterizing_2022}, although we note that the LoNS increase at 532 nm is rather small with increasing zenith distance, as can also be seen in figure \ref{fig:snr_filter_hamamatsu}.

To estimate the systematic uncertainties of this atmospheric modeling we used the radiance ratio of the \textit{SKYCALC} model to the FORS~1 spectra presented by \cite{noll_atmospheric_2012}. As described in their paper, we divided the radiance ratio curve into two sections, one for B+V optical bands and one for longer wavelengths. We then calculated the 68\% (1\,$\sigma$) containment levels and used these as the initial uncertainty levels from the \textit{SKYCALC} model. \cite{noll_atmospheric_2012} mentions larger uncertainties for airmass $>2$. Therefore, we assume that the model uncertainty for each wavelength scales linearly with the airmass. Further systematic uncertainties for the light guide efficiency, non-linearity of the readout, and mirror reflectivity were taken from \cite{aleksic_major_2016-II}.
The uncertainty of the photon-detection efficiency for the PMTs was taken from \cite{aleksic_performance_2012}. For the SiPMs we used the spread between the used sensors for the systematic uncertainty calculation.

\begin{figure}[tb]
    \centering
    \includegraphics[width=1\columnwidth]{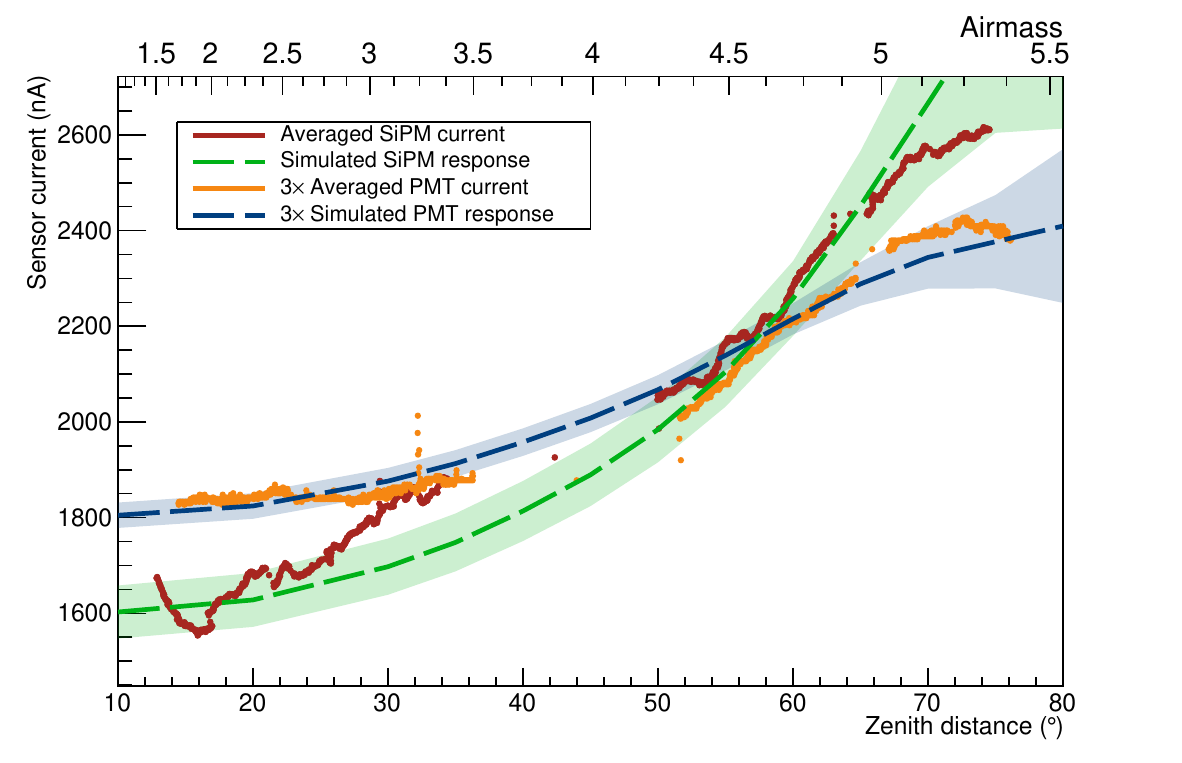}
    \caption{Average PMT (orange) scaled by a factor of 3 for better visibility and Hamamatsu SiPM DCs (red) compared to our predictions from our LoNS model based on \cite{benn_palma_1998} (blue, green). In both cases, the average of seven pixels is plotted during real telescope observations of one source during one night of moon-less data taking. The observations were all conducted after 23\,UT, which proves to be relevant due to a change of the street lighting at midnight local time as described in section 9 of \cite{fruck_characterizing_2022}. 
    }
    \label{fig:DC_vs_Zd}
\end{figure}

The measured direct current (DC) versus prediction from our model is shown in figure \ref{fig:DC_vs_Zd} where we show the average current of a PMT and a SiPM-based module during a moon-less night observing a single astrophysical target. We note that we accounted for the fact that the SiPM modules are located at the camera edge and therefore the observed zenith distance is slightly different for the SiPM pixels than for the central PMTs. Not accounting for this effect would, for example, lead to a 7.5\% error in the determined airmass at 70\degr. 
It can be seen that our LoNS model sufficiently describes the qualitative changes of the measured pixel current averages. The biggest discrepancy is seen for the SiPM pixels at very large zenith distances where the model over-predicts the measurement by 4\% at ZD=70\degr, 9\% at ZD=74\degr. The brightest star in the field of view (FoV) near the SiPM pixels during this observation had the magnitude $\sim6.4$. This star could be an explanation for the discrepancy between the LoNS model curve and data points around ZD=30\degr where it came closest to the SiPM pixels. There was no star brighter than magnitude 6.5 close to the relevant PMT pixels during the observation time. The growing difference between the LoNS model curve and SiPM current above $\mathrm{ZD}\sim70\degr$ is reflected by the diverging uncertainties due to the extrapolation uncertainties with the rapidly increasing airmass.

We draw $\pm 1 \sigma$ containment contours around the predicted model curve which increase linearly for airmass $>2$ similar to our ansatz for the atmospheric modeling above. We take the determined width of these containment contours as the systematic uncertainty of fast atmospheric changes and include it in our uncertainty calculation of the signal-to-noise ratio below.

Comparing the PDE curves point-wise multiplied with the LoNS spectrum provides an estimate on the relative increase of detected LoNS signal with respect to the PMTs: The Excelitas-based pixels will see $\sim3$ times, the Hamamatsu-based pixels $\sim4.9$ times
and the SensL based pixels $\sim4.4$ times more LoNS.
Based on the LoNS rate measured by \cite{bartko_tests_2005} and accounting for the hardware changes since their measurement we calculate a LoNS induced rate of about $550\,\mathrm{MHz}$ for the Excelitas pixels, $880\,\mathrm{MHz}$ for the Hamamatsu pixels and $800\,\mathrm{MHz}$ for the SensL pixels. The results are summarized in table \ref{tab:pixel_calc}. The uncertainties of the LoNS-induced rate were calculated using the uncertainty range from our LoNS model.\\
The FWHM of a SiPM single photoelectron (abbreviated as phe) is about $3\,\mathrm{ns}$. With such high background rates close to the order of $\mathrm{GHz}$ this results in a strong pileup of single phe which makes it impossible to distinguish between single phes during astrophysical observations.

\begin{table*}[htb]
\centering
\begin{tabular}{lllc}
\toprule
           & \begin{tabular}[c]{@{}l@{}}Efficiency on Cherenkov\\ wrt. MAGIC PMTs\end{tabular} & \begin{tabular}[c]{@{}l@{}}LoNS rate\\ sim.(MHz)\end{tabular} & \begin{tabular}[c]{@{}l@{}}LoNS rate\\ meas. (MHz)\end{tabular} \\ \midrule
MAGIC PMTs & 1                                                                                    & $180 \pm   24 $    & 232                                        \\
Excelitas  & $0.75 \pm 0.13_\mathrm{sys} $                                                        & $550 \pm   110$    &                                        \\
Hamamatsu  & $1.7\:\;  \pm 0.2_\mathrm{sys}  $                                                    & $880 \pm   150$    & 864                                        \\
SensL      & $1.6\:\;  \pm 0.2_\mathrm{sys}  $                                                    & $800 \pm   130$    &                                       \\
\bottomrule
\end{tabular}
\caption{Summary of estimated SiPM module performances on Cherenkov light from air showers at low zenith distances and LoNS photons. 
}
\label{tab:pixel_calc}
\end{table*}

The direct measurement of the LoNS rate is further complicated due to the sliding window peak search charge extraction algorithm used by MAGIC in the standard analysis \cite{aleksic_major_2016-II}.
This algorithm will only detect a maximum of one LoNS-induced pulse in the readout window and will ignore all other potentially occurring LoNS pulses.
To measure the LoNS rate we therefore integrate the charge in a fixed window in the center of the waveform of $\sim65000$ pedestal events recorded with an opened camera observing at a low zenith distance during a moon-less night. These pedestal events are recorded at a fixed frequency of 25\,Hz during observations \cite{magic_collaboration_major_2016}. Stars and atmospheric phenomena in the FoV can affect the measured LoNs. To be able to directly compare the PMT and SiPM-based pixels we used a data run during which the Hamamatsu SiPM module was installed in the camera center, thus being surrounded by PMT pixels. This provides us with 12 symmetrical next neighbors, compared to just 2 asymmetrically distributed neighbors when the SiPM modules are installed in their usual location at the camera edges (see figure \ref{fig:camera_front}). This data set is further described for the evaluation of the SiPM module performance on air shower Cherenkov light in section \ref{sec:cherenkov_light_performance}. The number of phe in the fixed window will show Poissonian fluctuations due to the LoNS at the level of $\sqrt{\mathit{LoNS}}$. With this method, we measure an averaged LoNS rate during this observation of 232\,MHz in the neighboring PMT pixels and 864\,MHz in the Hamamatsu SiPM pixels in the camera center. The determined Hamamatsu SiPM LoNS rate agrees very well with the calculation summarized in table \ref{tab:pixel_calc}. The MAGIC PMT LoNS rate is almost 30\% above the predicted value and outside the 1-$\sigma$ uncertainty range of the averaged value. This discrepancy is primarily caused by a single PMT pixel showing a rate of 442\,MHz. Excluding this outlier would reduce the determined PMT LoNS rate to 197\,MHz which is well within the uncertainty. The brightest star in the vicinity of the relevant PMT pixels was magnitude $> 8$ so we can exclude a star being the cause for this high rate. It is however conceivable that a misaligned mirror panel was reflecting some other star close to the FoV into this pixel.\\
Depending on the Moon phase the LoNS can significantly increase compared to a moon-less night. The spectrum of the moonlight is very similar to the black body approximating the solar spectrum, but slightly colder (less blue) due to the lunar albedo \cite{jones_advanced_2013}. The spectrum of the moonlight reaching the IACT camera depends on the angular separation between the observation target and the Moon \cite{ahnen_performance_2017}. Non-optimal atmospheric conditions due to the presence of, for example, dust or clouds can further increase the LoNS due to increased scattering of moonlight. While it is safe to operate a SiPM-based camera during high LoNS levels from a hardware point of view, linearity and temperature considerations (due to the flow of high currents) need to be carefully assessed. The use of a bandpass filter to suppress the LoNS is discussed in section \ref{sec:performance_filter}.

\subsection{Signal-to-noise ratio}
\label{sec:snr}

We define the signal-to-noise ratio (SNR) of signal $S$ and noise $N$ as the ratio of the mean generalized Poisson-distributed signal $\mu_{\mathrm{signal}}$ to the standard deviation of the generalized Poisson-distributed noise $\sigma_\mathrm{noise}$:\\
\begin{equation}
    \mathit{SNR} \coloneqq \frac{\mu_{\mathrm{signal}}}{\sigma_\mathrm{noise}} = \frac{\langle S \rangle }{\sqrt{\langle S \rangle  + \langle N \rangle}} \times \sqrt{1-\lambda} 
\label{eq:snr}
\end{equation}
with $\sqrt{1-\lambda}$ and $\lambda = \ln{\frac{1}{1-p}}$ with $p$ the cross-talk probability describing the additional noise contribution from a cross-talk to the mean Poisson-distributed signal $\langle S \rangle$ and noise $\langle N \rangle$. The effects of cross-talk on the expectation value and variance of the signal and noise, described by a branching Poisson process, are derived in \cite{vinogradov_analytical_2012}. General properties of the generalized Poisson distribution are discussed in \cite{consul_interesting_1973}.

In a real measurement, $S$ always consists of the real signal $S_\mathrm{true}$ and the noise contribution underlying it: $\langle S \rangle = \langle S_\mathrm{true} + N \rangle = \langle S_\mathrm{true} \rangle + \langle N \rangle$. Therefore, the signal-to-noise ratio at trigger level $\mathit{SNR_\mathrm{trigger}}$ is
\begin{equation}
    \mathit{SNR_\mathrm{trigger}} = \frac{\langle S_\mathrm{true}\rangle}{\sqrt{\langle S_\mathrm{true} \rangle + 2\times\langle N \rangle}}  \times \sqrt{1-\lambda}.
\label{eq:snr_trigger}
\end{equation}

For a real measurement without background subtraction, one would get 
\begin{equation}
    \mathit{SNR_\mathrm{measurement}} = \frac{\langle S_\mathrm{true}\rangle + \langle N \rangle}{\sqrt{\langle S_\mathrm{true} \rangle + 2\times\langle N \rangle}}  \times \sqrt{1-\lambda}.
\label{eq:snr_measurement}
\end{equation}
This is equivalent to the formalism used for example by \cite{hamamatsu_snr_simulator} (see also \cite{konopelko_effectiveness_1999})  with explicitly including the cross-talk.
For the reasons given above additional excess noise factor (ENF) contributions as described in \cite{vinogradov_status_2020} adding to the noise term are very small and we will only consider the background noise caused by the LoNS.

\begin{figure}[tb]
\centering
\includegraphics[width=1\columnwidth]{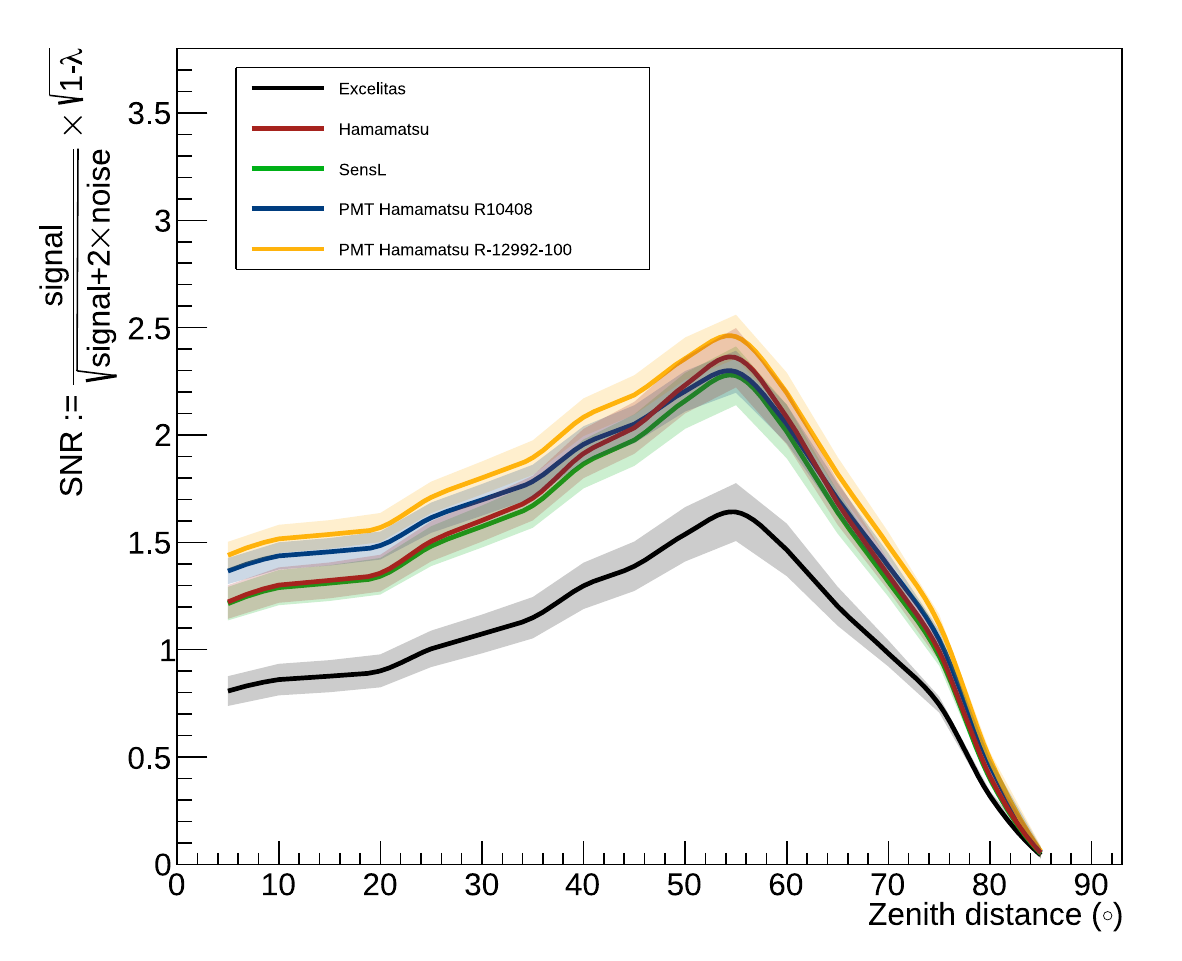}
\caption{SNR of a camera completely based on a given sensor for simulated 5\,TeV gamma-ray events for different zenith distances. The diminishing uncertainties at larger zenith distances are caused by the diminishing signal (see eq. \ref{eq:snr}).
}
\label{fig:SN_MC}
\end{figure}

The change of the SNR with increasing zenith distance is a combination of increasing atmospheric absorption, increasing LoNS background, and decreasing shower image size. We simulated 5\,TeV gamma-ray events at zenith distances between 5\degr and 85\degr. The simulated camera was fully equipped with one given sensor type.
The result is shown in figure \ref{fig:SN_MC} where we show the SNR using the photons in a single pixel. At low zenith distances PMTs show a slightly higher SNR. From medium to very large zenith distances, our SiPM-based pixels show a comparable SNR to the MAGIC PMTs. 
The newer PMTs used in LSTs show consistently higher SNR but agree with our second generation of SiPM-based pixels within uncertainties for ZD$>50\degr$.
It is interesting to note that even in a hypothetical case of absence of cross-talk in SiPMs ($p=0 \Rightarrow \lambda=0$), these only marginally outperform the PMTs in the range between $\sim45\degr$ to $\sim65\degr$ while showing comparable SNR everywhere else.

The calculation or comparison of SNRs does however not directly allow us to infer similar changes in a complete telescope system sensitivity. This would require full and detailed simulations for the trigger, analysis pipelines etc.\ which is beyond the scope of this study. A specific realization was studied in e.g.~\cite{arcaro_study_2022}.

\subsection{Linearity and cell recharging (gain stability)}

The finite number of Geiger-mode avalanche photodiode (G-APD) cells on a SiPM introduces some non-linearity in the response under certain conditions (see e.g.\ \cite{vinogradov_efficiency_2011, vinogradov_status_2020}).

The saturation effect that causes a non-linearity at high signal inputs can safely be neglected for the detection of atmospheric air showers, which are too faint to suffer noticeably from this effect. This holds true even taking into account the LoNS as we will discuss below.\\
At high background rates, G-APDs can be triggered before fully recovering from a previous discharge. This would increase the gain variance and lower the mean gain of the device. We calculated the time for a $90\%$ recharge $\tau_\mathrm{90}$ and the Poisson probabilities of another LoNS photon hitting the same cell within this time period $p_\mathrm{90}$. We find $p_\mathrm{90} \leq 1.2\%$ 
for all three pixel designs. Here we neglect the fact that a G-APD can not develop another charge avalanche right away but needs to be charged to a minimum level. The presented value can therefore be regarded as a conservative estimate.

Under the influence of strong moonlight $p_\mathrm{90}$ increases significantly. For simplicity, we assume here that the LoNS rate increases by a factor of 12, which is the standard operational upper limit for MAGIC (see \cite{ahnen_performance_2017}) since the exact conditions strongly depend on the telescope's orientation, angular distance to the Moon (due to Rayleigh scattering) and the Moon phase. We calculate $p_\mathrm{90}$ to be $\sim 13\%$ for the Excelitas-based module, $\sim 1.2\%$ for the Hamamatsu-based module, and $\sim 0.7\%$ for the SensL-based module. This, especially for the first-generation module, will result in reduced charge resolution and increased systematic uncertainties for observations under strong moonlight.\\
Table \ref{tab:pixel_cell_recharge} summarizes these calculated probabilities and connected parameters.

\begin{table*}[th]
\centering
\begin{tabular}{lllllll}
\toprule
          & \thead{Cells/SiPM} & \thead{Cells/pixel} & \thead{Accessible} & \thead{$\tau_{90}$ (ns)} & \multicolumn{2}{c}{\thead{$p_{90}$ (\%)}} \\
          &                             &                              &        \thead{active area (\%)}                                                                            &                                                                             & \thead{dark}                                 & \thead{moon}                                 \\
          \midrule
Excelitas & 14,400                                                     & 100,800                                                     & 69                                             & 219                                       &                  1.2                                         & 13                                                           \\
Hamamatsu & 6,400                                                      & 57,600                                                      & 84                                                                    &      69                                                      &                     0.1                                      & 1.2                                                          \\
SensL     & 22,300                                                     & 200,600                                                     & 87                                                                    &       111                                                     &                         0.1                                   & 0.7                                                         \\
\bottomrule
\end{tabular}
\caption{Characteristic parameters for the G-APD cell recharging under the influence of LoNS.
}
\label{tab:pixel_cell_recharge}
\end{table*}

\subsection{Thermal management and temperature stability}
\label{sec:thermal}

\begin{figure}[htb]
\centering
\includegraphics[width=1\columnwidth]{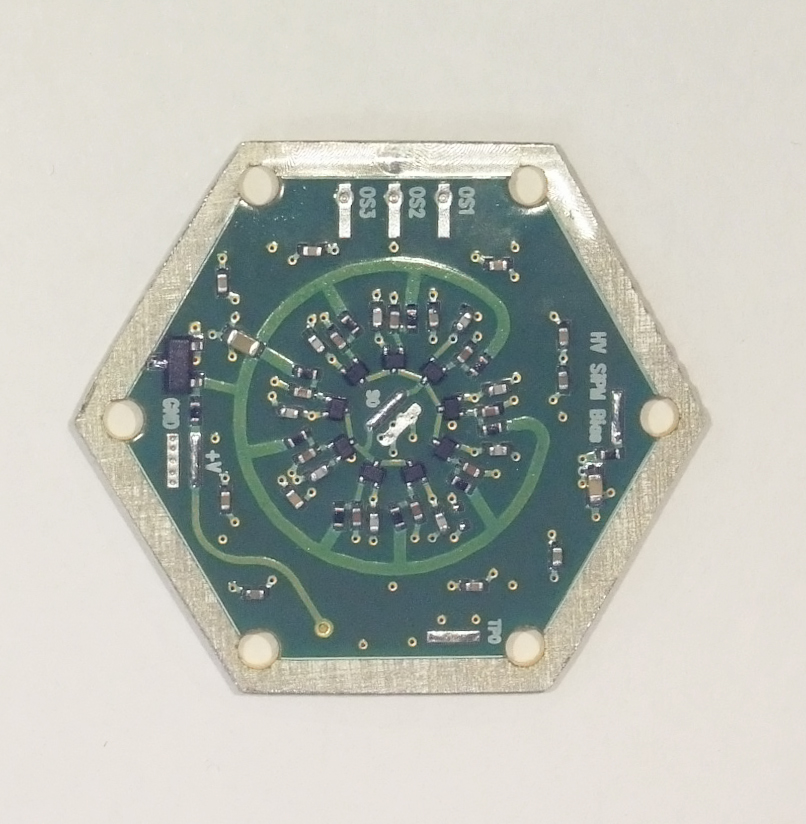}
\caption{Bottom view of a second generation SiPM pixel. The border of the central aluminum core layer can be seen surrounding the slightly smaller bottom PCB layer. The aluminum core is in thermal contact with the metal structure of the SiPM module, conducting the dissipated heat from the electronics components to the cooling plates further back in the camera housing.  
}
\label{fig:pixel_alu_core}
\end{figure}

\begin{figure}[htb]
\centering
\includegraphics[width=1\columnwidth]{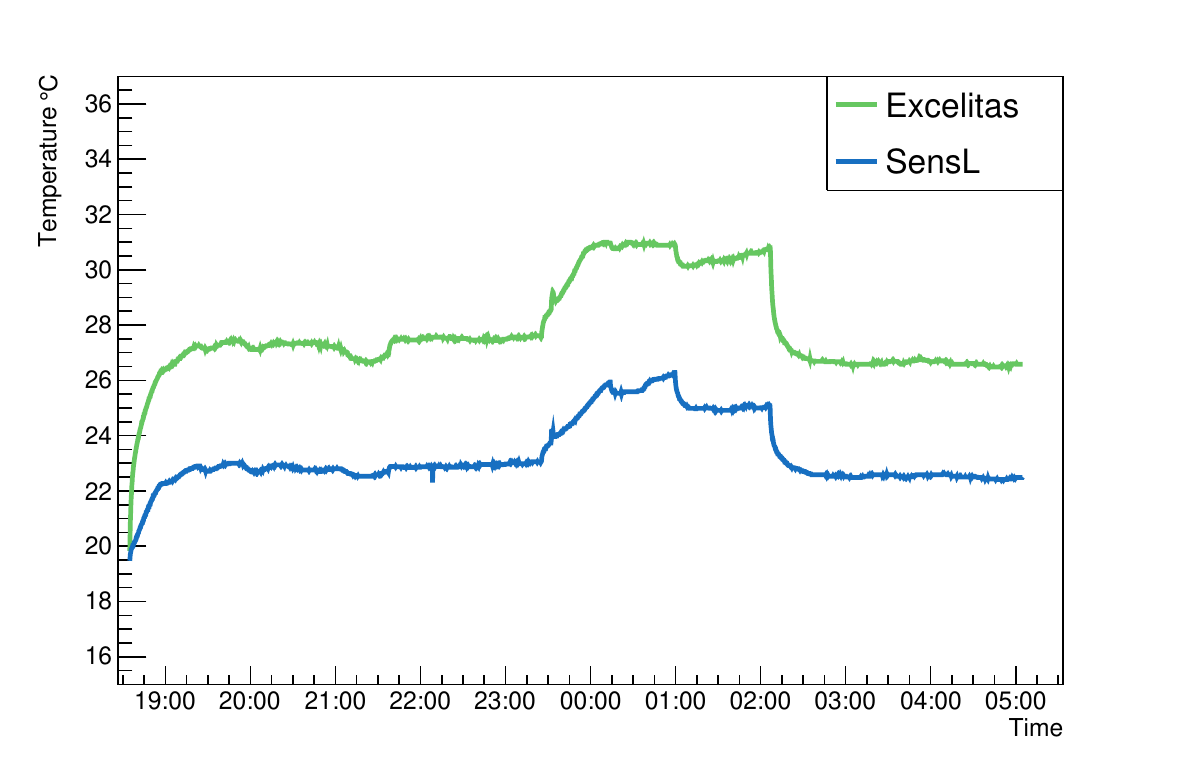}
\caption{Mean pixel temperature of the seven pixels of a first and a second-generation SiPM-based module during a dark night, moon observations, and an engineering run. Figure taken from \cite{hahn_performance_2023}.}
\label{fig:pixel_temperature}
\end{figure}

The first SiPM-based detector module used regular multi-layer PCBs for its pixels. The heat generated in those pixels flows along the pixels, through the PCB substrate, to the aluminum front structure where these are mounted. The heat is then guided along the two inner and six outer metal rods $\sim14\,\mathrm{cm}$ to the camera cooling and temperature stabilization plate.\\
For the second generation of SiPM-based modules, we used multi-layer aluminum core PCBs and an improved inner mechanical module structure. Figure \ref{fig:pixel_alu_core} shows the bottom side of such a pixel PCB. One can see the inner aluminum core, surrounding the smaller bottom PCB layer. This electrically insulated inner aluminum layer of the PCB is in thermal contact with the module's front aluminum structure, directly transferring the heat generated in the pixels. The module's inner two metal rods were replaced by a single thickened plate, mounted with thermal pads, further improving the thermal conductivity between the pixels and the camera cooling plate. We tested for potential broadening of the SiPM waveform due to possible capacitive coupling between the SiPM and the comparatively thick inner aluminum layer but found none.\\
The mean pixel temperatures of the seven pixels of a first and a second-generation SiPM-based module during regular telescope operations are shown in figure \ref{fig:pixel_temperature}. After the regular electronics warm-up, the SiPM pixels were operated at a stable temperature until $\sim$23:30 when the bright Moon (83\% moon illumination, \nth{4} night after full Moon) starts to rise. With the increasing background light, the temperature of both pixel types increases until a new stable plateau is reached. The plateau level depends on the angular separation between the observed astrophysical source and the Moon, as can be seen from the step at $\sim$01:00 caused by the repointing of the telescope to a different celestial position. At $\sim$02:00 the camera is switched off for an engineering run and the pixels return to the same temperature as during the data taking under dark conditions. 

\section{Installation}
\label{sec:filed_tests}

\begin{figure}[tb]
\centering
\includegraphics[width=0.99\columnwidth]{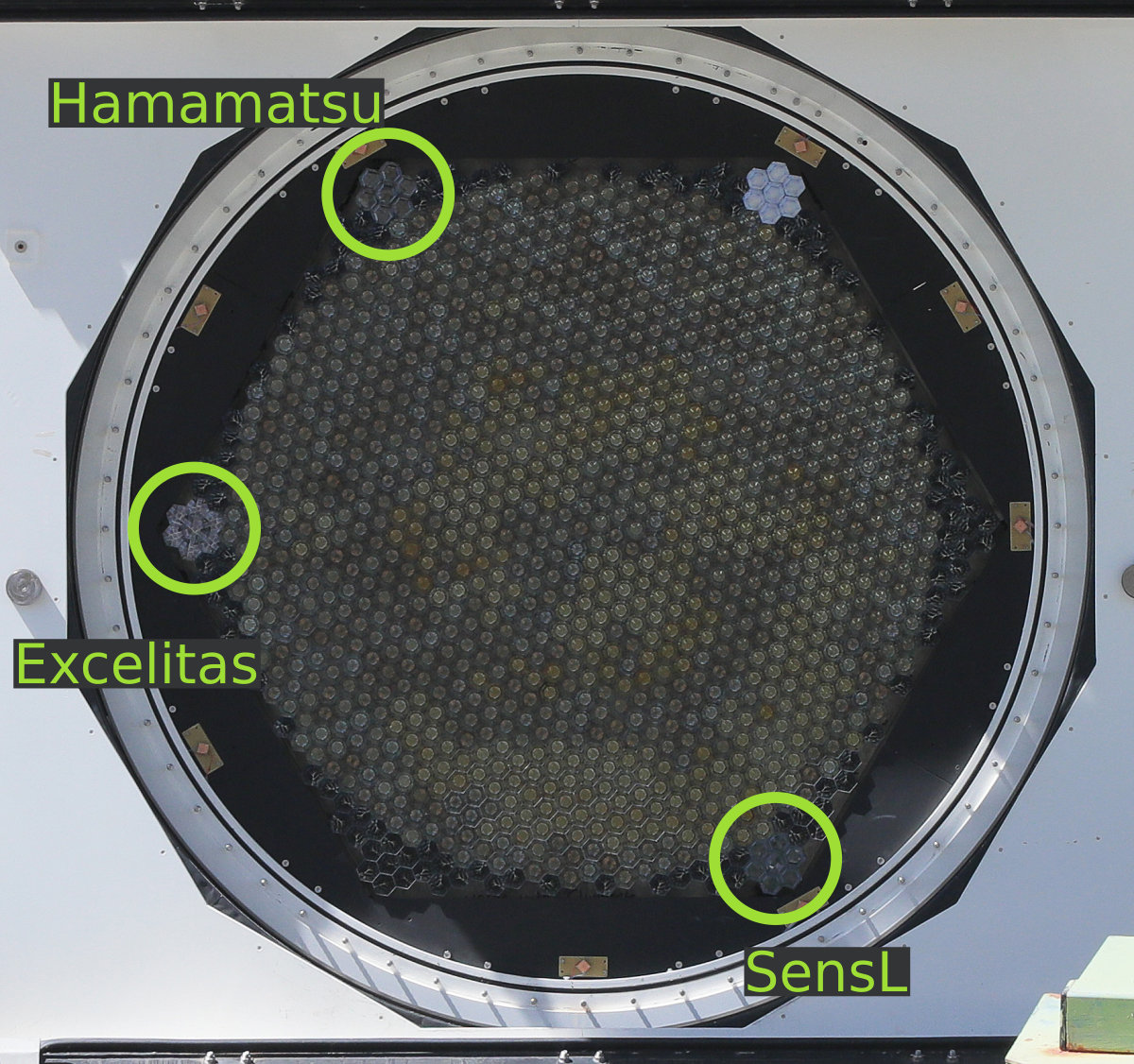}
\caption{Front view of the MAGICI-I camera with the installed SiPM prototype modules. The three SiPM-based modules described in this work are labeled. In addition, one can spot the installed light-trap module in the top right corner and another SensL SiPM-based module featuring an alternative light-guide design in the far right corner. See \cite{ward_light-trap:_2016, guberman_light-trap:_2019} and \cite{nasr_esfahani_effectivity_2019} for more details on these additional SiPM-based modules.
}
\label{fig:camera_front}
\end{figure}

The inner structure of the MAGIC camera housing which can be equipped with photodetectors is hexagonal. It is equipped with 1039 PMTs in a circular shape \cite{magic_collaboration_major_2016}. This leaves the six vertices of the hexagon open for the installation of prototype modules. A front view photo can be seen in figure \ref{fig:camera_front}.\\
We installed our first SiPM-based detector module in 2015 and the two second-generation SiPM modules in 2017. The Excelitas-based module is located in the left corner, the SensL-based module in the bottom right corner, and the Hamamatsu-based module in the top left corner. In the top right corner, one can also see a different SiPM-based prototype, the so-called light-trap which is discussed in \cite{ward_light-trap:_2016,guberman_light-trap:_2019}. The SiPM modules are controlled and monitored by the general MAGIC camera slow control. They are integrated into the MAGIC data-taking system though they are not used in the astrophysical data analysis. The SiPM modules record events in "parasitic" trigger mode, i.e.\ when the inner region of the PMT-based camera (see \cite{magic_collaboration_major_2016}) is triggered. The SiPMs are operated every night when MAGIC is taking gamma-ray data.

As mentioned in section \ref{sec:operational_conditions}, the analog optical signals are transmitted to the data acquisition system in the counting house via optical fibers. Due to the lower signal transit time of SiPMs compared to PMTs, we added an additional 1.5\,m of optical fiber to ensure that SiPM and PMT pulses arrive at about the same time.

\section{Calibration}
\subsection{Single photo-electron spectrum method}
\label{subsec:single_phe_calib}

\begin{figure}[tb]
\centering
\includegraphics[width=1\columnwidth]{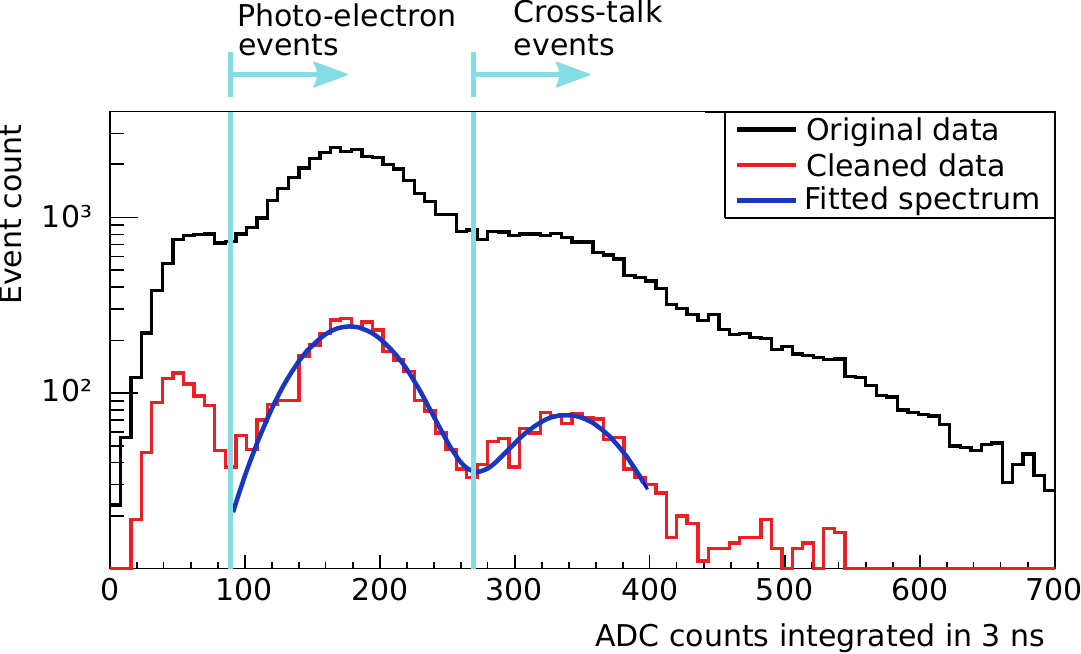}
\caption{Raw and filtered charge histogram of pedestal events. The range of single and the two photo-electron peak are fitted by a spectrum function. Figure taken from \cite{hahn_development_2018}.}
\label{fig:charge_histo}
\end{figure}

The PMTs of the MAGIC cameras are calibrated utilizing the detector's intrinsic F-factor, sometimes also known as ENF \cite{mirzoyan_calibration_1997, magic_collaboration_major_2016}. However, this method is not straightforwardly applicable for the calibration of SiPMs because of the fundamentally different origin of the excess noise in SiPMs \cite{renker_advances_2009}. Therefore we use the pulse charge spectrum of single photo-electrons for the calibration of gain and the determination of the cross-talk probability when possible. An alternative method, based on the F-factor is described in section \ref{subsec:ffactor}.\\
As explained above in section \ref{sec:expectations_spectra} the pileup renders it impossible to distinguish the individual photo–electron peaks in
a charge histogram with the camera illuminated by the LoNS. But with the camera lids closed the recorded events consist only of pedestals and dark counts with a probabilistic cross-talk contribution. $10^{5}$ of such events are recorded at a fixed frequency prior to data taking every night.\\
The signal charge is extracted with a sliding window integrating peak search algorithm, the same procedure as used by MAGIC \cite{aleksic_major_2016-II}. For a more detailed description and comparison with other charge extractors, we refer to \cite{albert_fadc_2008}. The total length of the readout window for an event is $30\,\mathrm{ns}$ but we exclude the first and last $3\,\mathrm{ns}$ from the charge integration to avoid pulses which are only partially contained in the readout window or suffer from additional noise near the readout window edge \cite{biland_calibration_2014}. A baseline is calculated from samples outside the integrating window and subtracted from the extracted charge. The so extracted charge of this $3\,\mathrm{ns}$ wide sliding window suffers from a positive bias because the extractor intrinsically favors positive fluctuations.
Therefore, the pedestal peak of the charge histogram is not Gaussian distributed and not centered around zero but shifted and skewed towards higher values. This shift and skew are decreasing with the number of photo-electrons because of a decreasing noise contribution to the signal but they are already sufficiently small even for single phe pulses. Therefore the distance between the single and the two photo-electron peaks in the charge histogram provides the calibration constant to convert the recorded charge into number of photo-electrons. The width of $3\,\mathrm{ns}$ of the charge extraction window is on the order of the FWHM of the summed SiPM pulses, see section \ref{sec:electronics_summation}.\\
The individual peaks of the charge histogram are not well separated as can be seen in figure \ref{fig:charge_histo}. Fitting this charge spectrum would give less reliable results with large uncertainties. We developed an event filtering algorithm to improve the charge separation and consequently the calibration. The filtering conditions are:
\begin{itemize}
    \item the pulse must be approximately Gaussian in shape,
    \item the pulse FWHM must be narrow enough to exclude overlapping pulses,
    \item the baseline throughout the full readout window must be stable,
    \item and the readout window must not contain after pulsing or delayed cross-talk events.
\end{itemize}
About $5\%$ of the initially recorded events fulfill these restrictive conditions.
The resulting histogram, which shows a significantly improved charge separation, is also plotted in figure \ref{fig:charge_histo}. This filtered histogram is then fitted to obtain the calibration constant. By integrating the number of events in the unfiltered histogram with two or more photo-electrons and dividing it by the number of events with at least one phe we calculate the cross-talk probability. The integration ranges are determined from the filtered data but to avoid biases from the described event filter the unfiltered data has to be integrated. The probability of two dark count events occurring so close in time to be interpreted as a cross-talk event is negligible.

\subsection{SiPM F-factor and the difference to the PMT F-factor method}
\label{subsec:ffactor}

In case the charge separation is not sufficient and prevents a reliable fit of the individual peaks in the charge histogram one can use an adapted F-factor based calibration similar to the PMT calibration method. For PMTs the main contribution to the excess noise is variations in the multiplication on the first dynode \cite{zworykin_secondary_1936, shockley_theory_1938} and the inelastic back-scattering of accelerated electrons off the dynodes. For SiPMs cross-talk is the cause of an excess noise factor $F > 1$ \cite{renker_advances_2009}. Still, the general relation between mean charge $\overline{Q}$, the mean charge distribution's $\mathit{FWHM}$, excess noise $F$ and mean number of detected photo-electrons $\overline{n_\mathrm{phe}}$ from e.g.\  \cite{teich_excess_1986} or \cite{wright_monte_1987} holds and can be transformed to:

\begin{equation}
\overline{n_\mathrm{phe}} = 8 \cdot \ln{2} \cdot F^{2} \cdot \left( \frac{\overline{Q}}{\mathit{FWHM}} \right) ^{2}.
\label{eq:ffactor_gen}
\end{equation}

\cite{vinogradov_analytical_2012} showed that a branching Poisson process provides expressions for the probability distributions and excess noise of SiPM signals with cross-talk. We adopt

\begin{equation}
F = \frac{1}{1-\lambda} = \frac{1}{1-\ln{\frac{1}{1-p}}}.
\label{eq:ffactor_crosstalk}
\end{equation}

This is equal to the $\mathit{ENF}$ of a branching Poisson process in \cite{vinogradov_analytical_2012} table 2 without the Taylor expansion. This analytical form proved to be computationally less demanding than calculating all necessary leading order terms of the Taylor expanded form.\\
The cross-talk probability $p$ depends on the applied over-voltage $U_\mathrm{over}$. Also the dark current $I_\mathrm{dark}$ rises with increasing over-voltage. We conducted lab measurements of $p$ and $I_\mathrm{dark}$ at varying $U_\mathrm{over}$. To achieve transferable results we read out $I_\mathrm{dark}$ and $U_\mathrm{over}$ with a setup identical to the slow control electronics in our SiPM modules. By using $p(I_\mathrm{dark})$ the dependence on $U_\mathrm{over}$ is removed
and we can directly calculate $p$ and consequently $F$ for every given $I_\mathrm{dark}$ in a meaningful range. Without the dependency on $U_\mathrm{over}$ also the temperature dependency is removed for our operating conditions. For the calibration of a SiPM module during telescope operations, we use the measured $I_\mathrm{dark}$ in parallel to the dark data run for the single photo-electron calibration method with closed camera lids.\\
It has to be emphasized that in contrast to the first method this excess-noise-based calibration method relies on lab measurements of $F$, similar to the calibration of PMTs \cite{magic_collaboration_major_2016}. \\
$\overline{Q}$ and $\mathit{FWHM}$ are extracted from calibration events, records of homogeneous light flashes of fixed brightness, timing, and wavelength. These calibration events are generated by the so-called calibration box, a laser flasher installed in the center of the mirror dish. A detailed description of the calibration box can be found in \cite{schweizer_optical_2002} with later upgrades, most importantly the improved homogeneity described in \cite{magic_collaboration_major_2016}.\\
We remark that an F-factor calibration method was used in the MC study of \cite{arcaro_study_2022} to simulate the calibration of SiPM pixels alongside PMT pixels.

\subsection{Long-term stability}

\begin{figure}[htb]
\centering
\includegraphics[width=1\columnwidth]{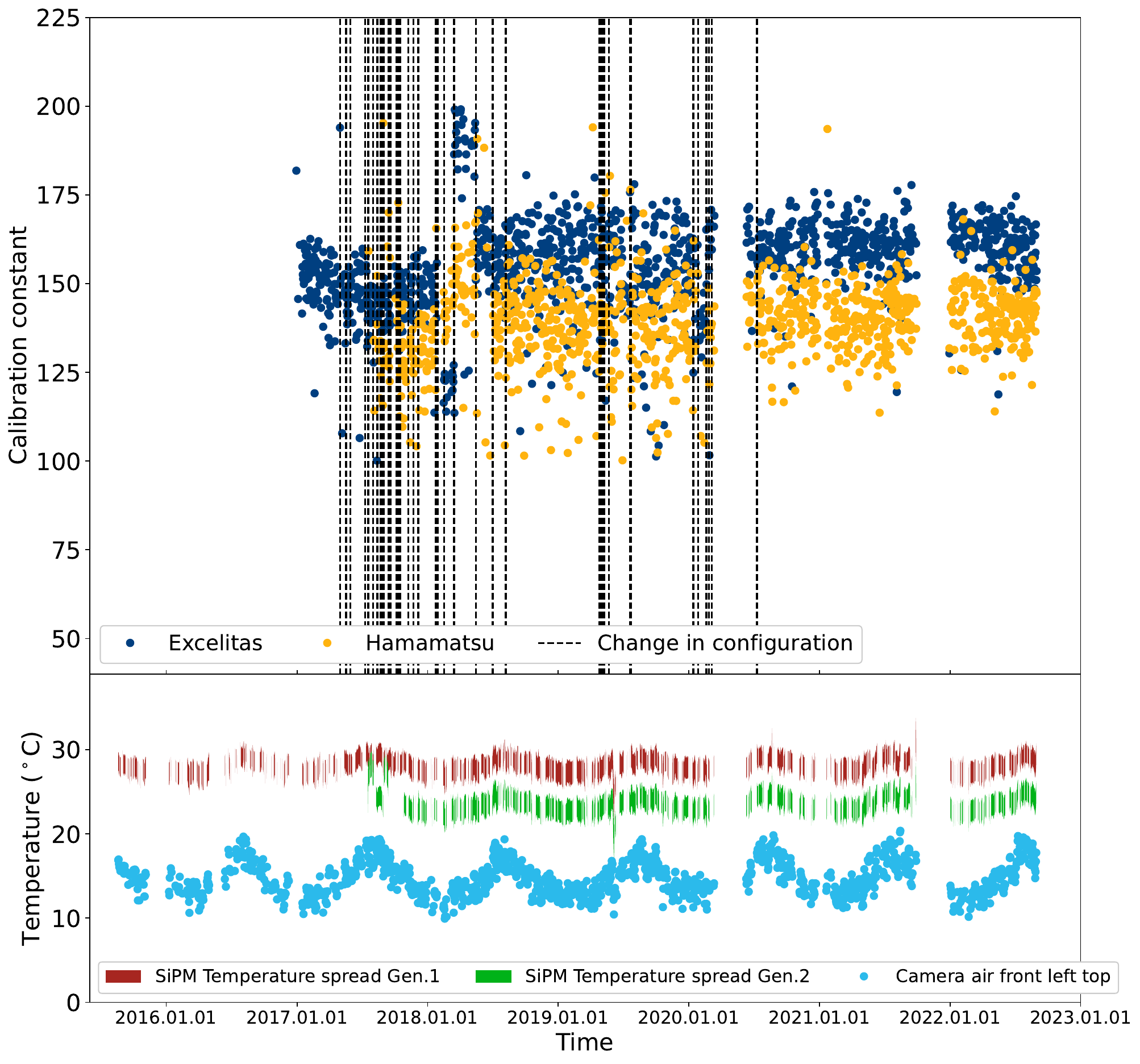}
\caption{Top: Calibration constants of an Excelitas and a Hamamatsu pixel since their installation. Vertical dashed lines indicate a change in the configuration e.g. a different voltage setting, connecting the pixel to a different readout channel, or the installation or removal of a band-pass filter. Bottom: SiPM pixel temperature spread during dark night times and seven pixels of one detector module of the given type (red and green). Camera air temperature measured at the front (blue). Two periods of interrupted operation can be seen due to stopped operations during the COVID-19 pandemic in 2020 and a volcanic eruption in La Palma end of 2021.
}
\label{fig:calib_stability}
\end{figure}

The SiPM modules are operated in parallel with the PMT-based scientific camera on a nightly basis. With several years of data collected so far, we studied the stability of the calibration. In figure \ref{fig:calib_stability} we show the calibration constant between 2017 and 2023. One can see seasonal oscillations of the calibration constants due to changing temperatures. This is matched by the behavior of the SiPM pixel temperatures and the air temperature inside the camera measured at the front, shown in the bottom panel of figure \ref{fig:calib_stability}. It is clearly visible that the air temperature variations are much larger than the pixel and module temperature variations. Apart from these temperature-related changes of the calibration constant, we see no long-term deterioration trend that would indicate an ageing of the SiPMs. Two larger gaps in the operation can be seen which are caused by the COVID-19 pandemic and a volcanic eruption.\\
We marked changes in the SiPM module configuration e.g. the change of the operational voltage or the installation of a bandpass filter. To study the differences between the readout channels of one module we permuted the optical fibers of the connected pixels during the course of two years. We however note that such differences are not part of the systematic uncertainty due to the pixel-wise calibration procedure. Only if a mean or global calibration procedure would be applied such a systematic uncertainty would need to be considered.

To assess the long-term systematic uncertainty of the calibration we binned the obtained calibration constants by pixel temperature during the calibration run and determined the standard deviation of the calibration constants within a 0.5\degC wide bin. We only used data that had the same hardware and software configurations to eliminate other effects from this study. The maximum long-term standard deviation obtained is $19\%$.\\
Finally, we would like to note that other calibration methods, which can even work with unresolved single-photoelectron spectra \cite{vinogradov_skewness-based_2022}, exist but were beyond the scope of this work and should be studied in a follow-up work.

\section{Performance}

\subsection{Calibration light performance}

\begin{figure}[tb]
\centering
\begin{subfigure}[b]{0.92\columnwidth}
    \centering
    \includegraphics[width=\textwidth]{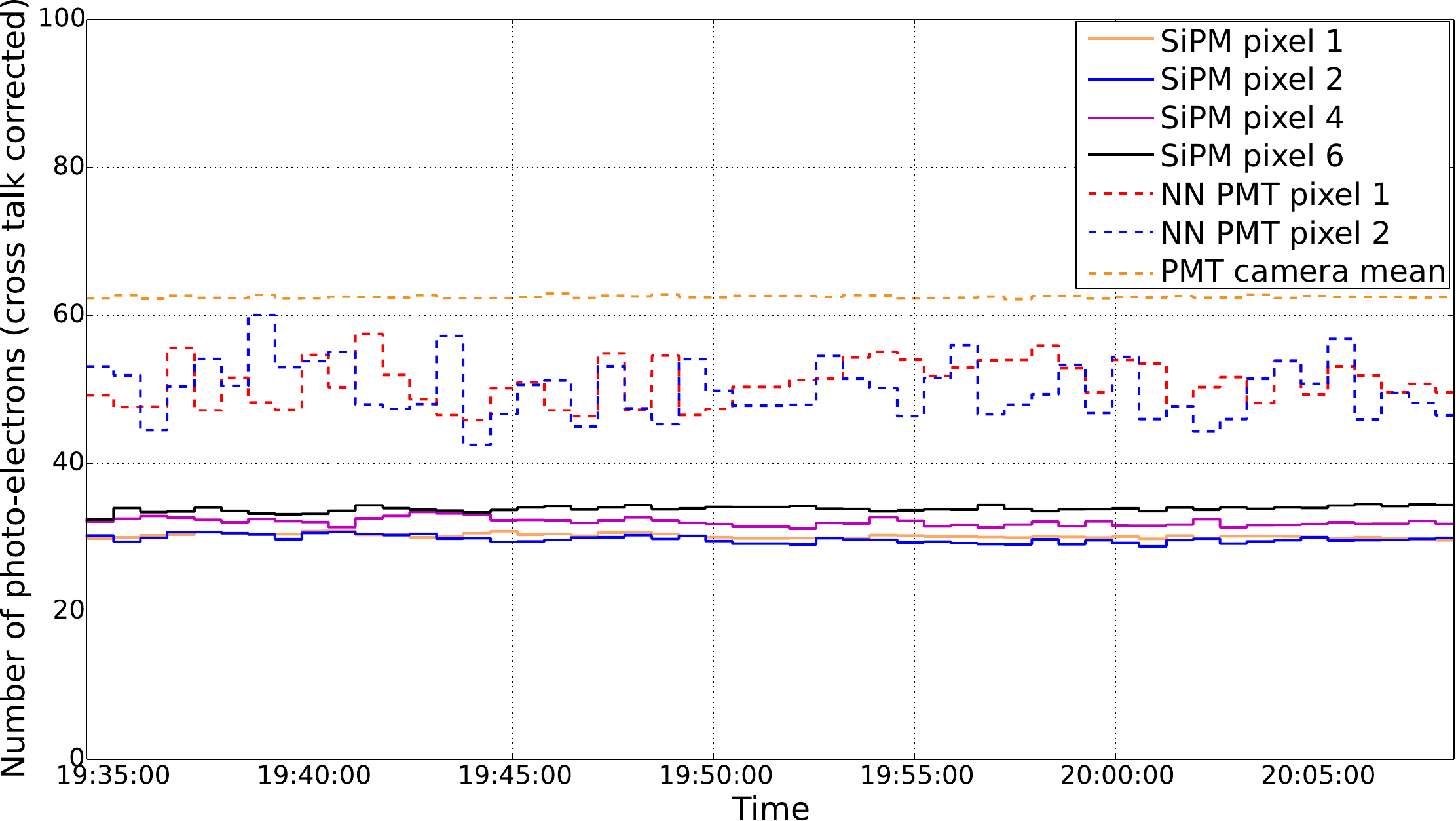}
    \caption{}
    \label{fig:nphe_calib_e}  
\end{subfigure}
\begin{subfigure}[b]{0.9\columnwidth}
    \centering
    \includegraphics[width=\textwidth]{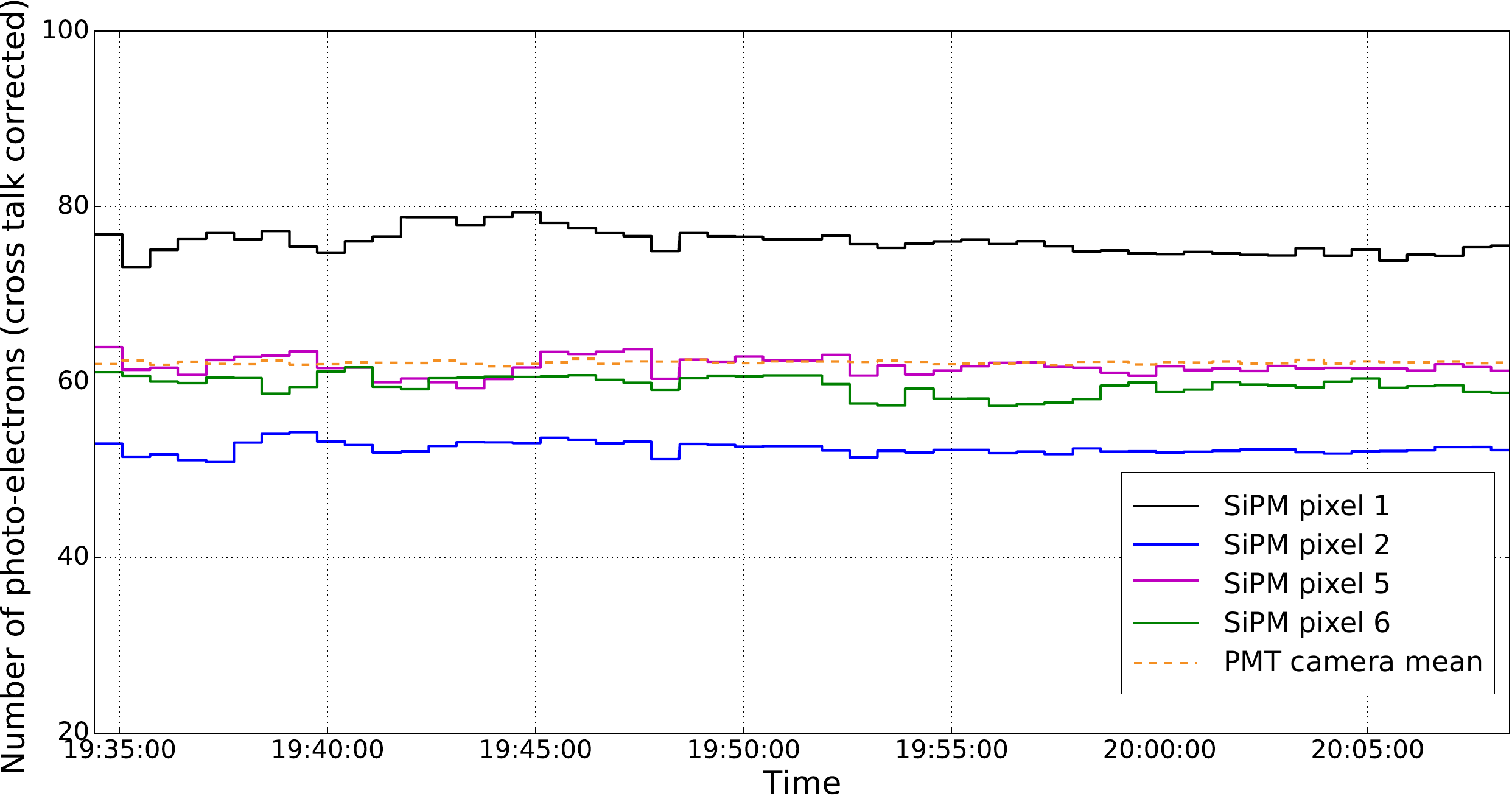}
    \caption{}
    \label{fig:nphe_calib_h} 
\end{subfigure}
\begin{subfigure}[b]{0.92\columnwidth}
  \centering  
    \includegraphics[width=\textwidth]{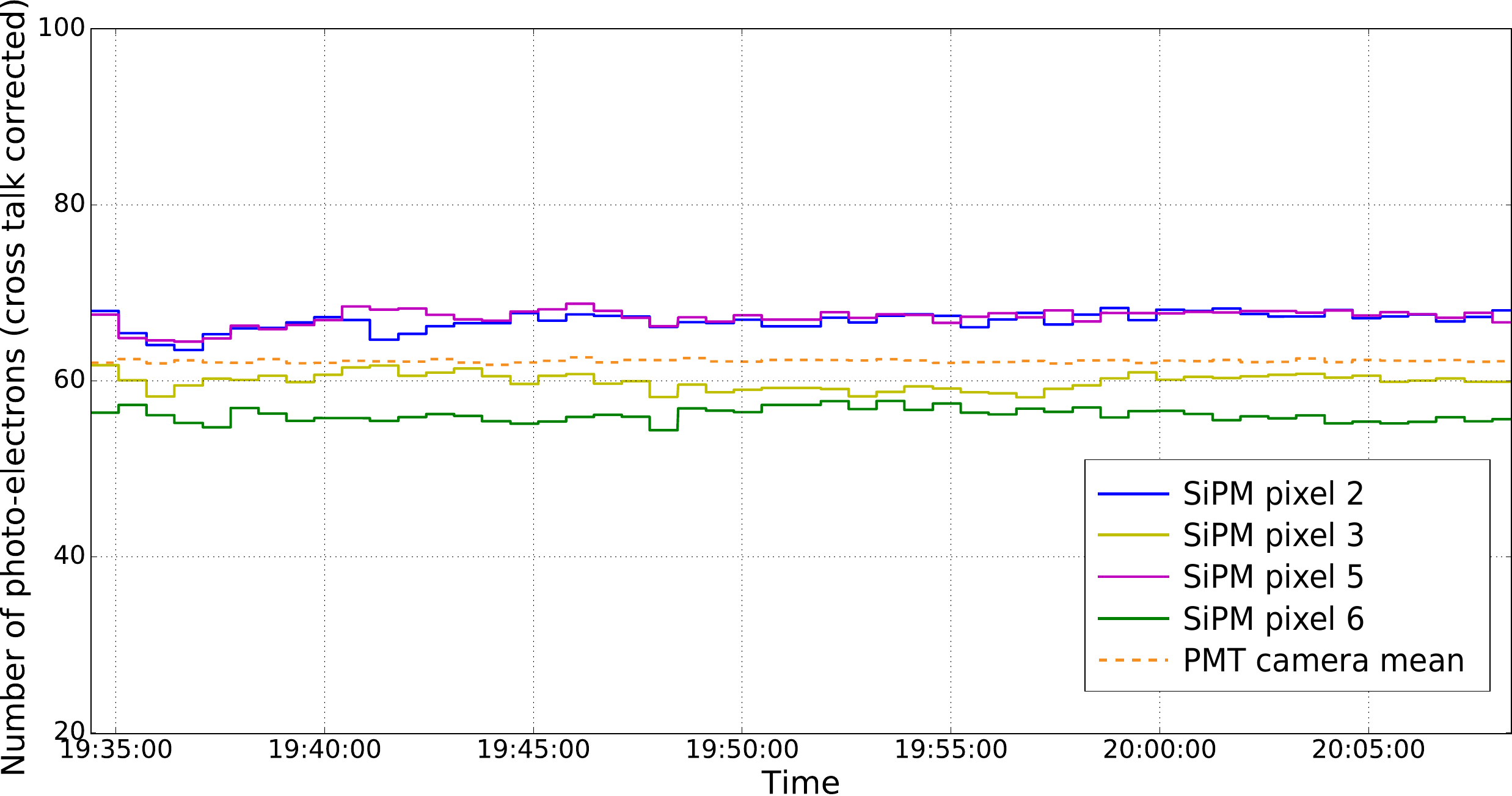}
    \caption{}
    \label{fig:nphe_calib_s} 
\end{subfigure}
\caption{Number of photo-electrons from calibration laser light flashes during the observation of one astrophysical source for the Excelitas (a), Hamamatsu (b), SensL (c) SiPM pixels, as well as the two next-neighbour PMT pixels and the PMT camera average. (b) and (c) from \cite{hahn_results_2019}.}
\label{fig:nphe_calib}
\end{figure}

First, we evaluate the performance of the SiPM-based modules with light pulses from the calibration laser. During data acquisition, the MAGIC cameras are homogeneously illuminated with the light pulses from a Nd:YAG laser, operating at the third harmonic (355\,nm). The duration of an individual pulse is 1\,ns, and the laser is fired at the frequency of 25\,Hz \cite{magic_collaboration_major_2016}. The Excelitas and the Hamamatsu-based SiPM modules were calibrated using the single photoelectron spectra, while the SensL-based SiPM module, due to the high dark rate,  was calibrated using the F-factor method. The mean extracted charge of 1000 calibration events ($\equiv40$\,s of observation) is shown in figure \ref{fig:nphe_calib} for the typical duration of an astrophysical source during one night. The obtained number of photo-electrons is in good agreement with the expectation using the PDE value of the SiPMs at the calibration laser wavelength from figure \ref{fig:PDE_LoNS_Cher}. In the MAGIC camera, the PMT pixel calibration factors are updated periodically using the same 1000 calibration events. A similar procedure can be applied also to the SiPM-based pixels, which are operated at the close to saturation applied overvoltage.

\subsection{Time resolution}
\label{sec:time_resolution}

\begin{figure}[tb]
\centering
\includegraphics[width=1\columnwidth]{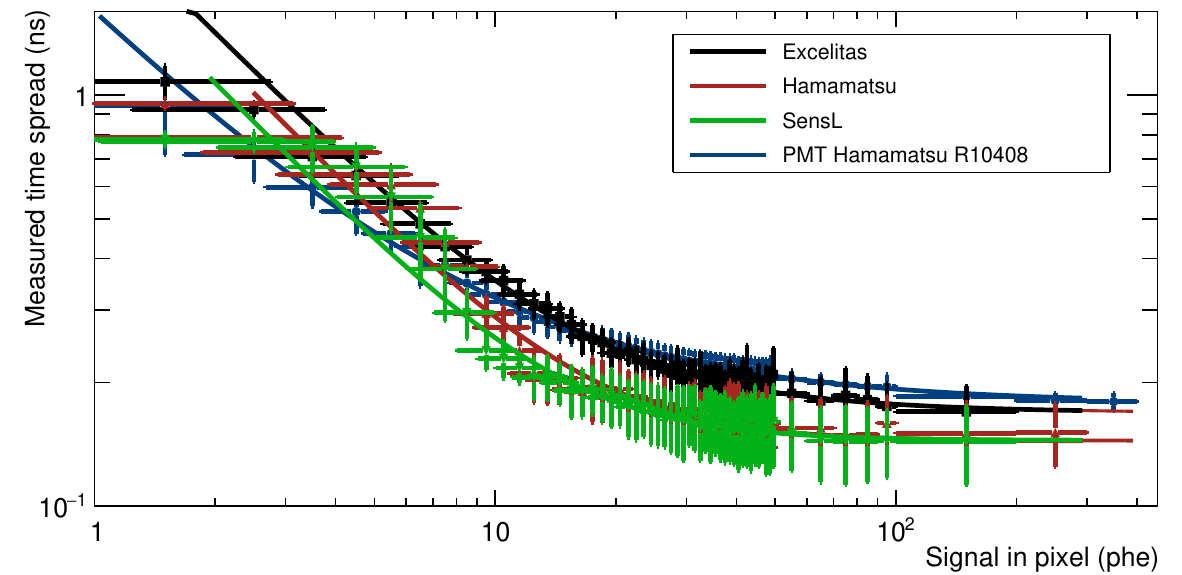}
\caption{Time spread of calibration light events of different detector module types measured from different calibration light intensities.}
\label{fig:time_resolution}
\end{figure}

To study the time resolution of the camera modules we followed the procedure in \cite{sitarek_analysis_2013}.
This required correcting for DRS4 non-equal sampling delay effects, which is part of the standard MAGIC analysis procedure \cite{magic_collaboration_major_2016} and also described in \cite{sitarek_analysis_2013}.
We first determine the mean arrival time for all light sensors of one type. Then we calculate the difference of the arrival time in each pixel to its type's mean arrival time. For 5000 events this arrival time deviation is calculated and subsequently fit by a Gaussian distribution. The final time spread $\Delta T$ is calculated as the mean of all pixel standard deviations. This procedure is repeated for calibration light flashes of several different intensities that yield from a few to several hundred photoelectrons. The number of photoelectrons $N_\mathrm{phe}$ is determined individually for each pixel and event. Due to the Poissonian distribution of the intensity of laser light pulses this binning in single phes might provide a more accurate result than the binning per filter setting as used in \cite{sitarek_analysis_2013}. To have sufficient events per bin we chose 10\,phe wide bins for signals $>$ 50\,phe and 100\,phe wide bins for signals $>$ 100 \,phe. The data was gathered while tracking a source-free region at 22\arcdeg zenith distance.\\
To study the individual contributions we fit the obtained time spread $\Delta T\left(N_\mathrm{phe}\right)$ with the parameterized form of individual time resolutions
\begin{equation}
    \Delta T\left(N_\mathrm{phe}\right) = \sqrt{\frac{T_0^2}{N_\mathrm{phe}} + \left(\frac{T_1}{N_\mathrm{phe}}\right)^2 + T_2^2}
\end{equation}
\label{eq:time_resolution}
according to \cite{albert_fadc_2008}. $T_0$ contains the contributions from the intrinsic pulse width and the transit time spread (TTS) of the pixel. $T_1$ is dependent on the pulse shape and resolution of the signal extraction. $T_2$ contains the residual ADC clock jitter and the possible jitter of other electronic components. Details can be found in \cite{albert_fadc_2008}.\\
The uncertainties of each data point are calculated from the Gaussian fit to the individual pixel arrival time deviations and the standard deviation of $N_\mathrm{phe}$. We add the square root of the expected number of LoNS-induced photoelectrons from section \ref{sec:expectations_spectra} in quadrature to the uncertainty of $N_\mathrm{phe}$, which becomes important for calibration light flashes of low intensity. The data and fit curves are shown in figure \ref{fig:time_resolution}.

To obtain the time resolution for extensive air showers $T_{0,\mathrm{EAS}}$ we correct the fit $T_0$ subtracting the intrinsic time spread of the calibration laser light pulse of 0.47\,ns and add the time spread of air shower photons in a single pixel of 0.3\,ns in quadrature. For the specific case of MAGIC-I, we also have to quadratically add an additional contribution due to the design of the MAGIC-I mirror dish consisting of staggered mirror tiles of 0.3\,ns.
The numerical values of these correcting factors were taken from \cite{sitarek_analysis_2013}. The time resolution parameters for extensive air showers are summarized in table \ref{tab:time_resolution_parameters}.

For the MAGIC PMTs we obtain results close to the parameters in \cite{sitarek_analysis_2013}. The rather small differences to the previously published values can be attributed to the decrease in the sampling frequency from 2.0\,GHz \cite{sitarek_analysis_2013} to 1.64\,GHz in November 2014 \cite{berti_study_2018}. The large uncertainties for the three SiPM's $T_0$ are caused by a combination of the large uncertainties in $N_\mathrm{phe}$ due to the higher sensitivity to LoNS and the poor sampling in the low $N_\mathrm{phe}$ region. The respective $T_0$ of the three SiPM-based modules is substantially below that of the PMT modules as can be expected from the intrinsically much smaller TTS of SiPMs. The resolution of the signal reconstruction is expected to be identical for SiPM and PMT modules, but the pulse shapes are different, resulting in a slightly larger $T_1$ for all three SiPM modules compared to the MAGIC PMTs. $T_2$ should not depend on the type of sensor, only the readout. In the fit we attribute the up to 0.03\,ns lower $T_2$ to variations in the different readout channels and the much lower number of SiPM pixels.
We note that fixing $T_2$ of the SiPM-based modules to the value obtained from the MAGIC PMT modules did not significantly alter the curvature of the fit and thus the best-fit values of $T_0$ and $T_1$, but mainly increased their uncertainties.\\
In summary, and independent of the small differences in $T_2$, the SiPM-based modules have better time resolution at more than 16 photoelectrons for the first generation and at more than 6 photoelectrons for the second generation SiPM modules. For a better comparison, one must consider the photon detection efficiency of the SiPM modules compared to the PMTs.
This means that for an equal illumination by air shower Cherenkov photons, the SiPMs of the second generation have a better time resolution across the whole range covered by this measurement.

\begin{table}[tb]
\begin{tabular}{cllll}
\toprule
          & \multicolumn{1}{c}{$T_{0,\mathrm{EAS}} (\mathrm{ns})$}  & \multicolumn{1}{c}{$T_1 (\mathrm{ns})$} & \multicolumn{1}{c}{$T_2 (\mathrm{ns})$}     &  \\ \midrule

PMT      & $0.74\pm0.01$  & $1.4\pm0.3$ & $0.178\pm0.005$ &  \\
Excelitas & $0.5\pm0.3$    & $2.7\pm0.4$ & $0.188\pm0.015$ &  \\
Hamamatsu & $0.2\pm0.5$    & $2.5\pm0.1$ & $0.144\pm0.005$ &  \\
SensL    & $0.2\pm0.4$    & $2.1\pm0.2$ & $0.145\pm0.006$ & \\

\bottomrule
\end{tabular}
\caption{Summary of obtained time resolution parameters in the MAGIC-I camera.}
\label{tab:time_resolution_parameters}
\end{table}

\subsection{Air shower Cherenkov light performance}
\label{sec:cherenkov_light_performance}

Due to the fact that the SiPM modules are installed at the edges of the imaging camera, they do not participate in the trigger as described in section \ref{sec:filed_tests}. To investigate their performance in the trigger, we installed the Hamamatsu SiPM module in the camera center for one observational night. The collected 40\,minutes of moon-less low-zenith data were analyzed by selecting only very large and bright shower images that illuminate the camera center. This corresponds to the 5\,TeV events we simulated in \ref{sec:expectations_spectra} section for the SNR calculations, which also illuminate most of the camera for low zenith distances.
The trigger and the image cleaning algorithms were the standard ones from MAGIC as described in \cite{lombardi_advanced_2011, aleksic_major_2016-II}. For the PMT pixels, the image cleaning was performed with the setting 6\,phe for the core pixel thresholds and 3.5\,phe for boundary pixels, which are the default settings used by MAGIC for observations under dark conditions \cite{ahnen_performance_2017}. 
We refer the reader to \cite{aleksic_observations_2011, shayduk_new_2005} for a description of the two thresholds and the image cleaning using the sum of the next neighbors. For the Hamamatsu SiPM pixels, we used the fluctuations of the LoNS-induced charge, extracted by the sliding window peak search algorithm, to calculate the corresponding image cleaning levels. We used 11.6\,phe as the threshold for core pixels and 6.8\,phe for the boundary pixels. We note that this differs from the findings by \cite{arcaro_study_2022}, who determined very similar cleaning thresholds for PMT and SiPM pixels in their MC study. We also note that the better time resolution of SiPMs compared to PMTs (section \ref{sec:time_resolution}) could be used for an improved image cleaning with a SiPM-based imaging camera. In this study, most of our SiPM pixels have four PMT pixels as neighbors. Therefore we kept the thresholds for the allowed time differences between neighboring pixels the same as for the MAGIC PMT camera, as described in \cite{aleksic_major_2016-II}.
Very large shower events can saturate the individual pixels. We, therefore, exclude events that contain PMT pixels with more than 700\,phe or Hamamatsu SiPM pixels with more than 1500\,phe although we note that this affects only 5\,\% of events.

\begin{figure}
    \centering
    \includegraphics[width=1\columnwidth]{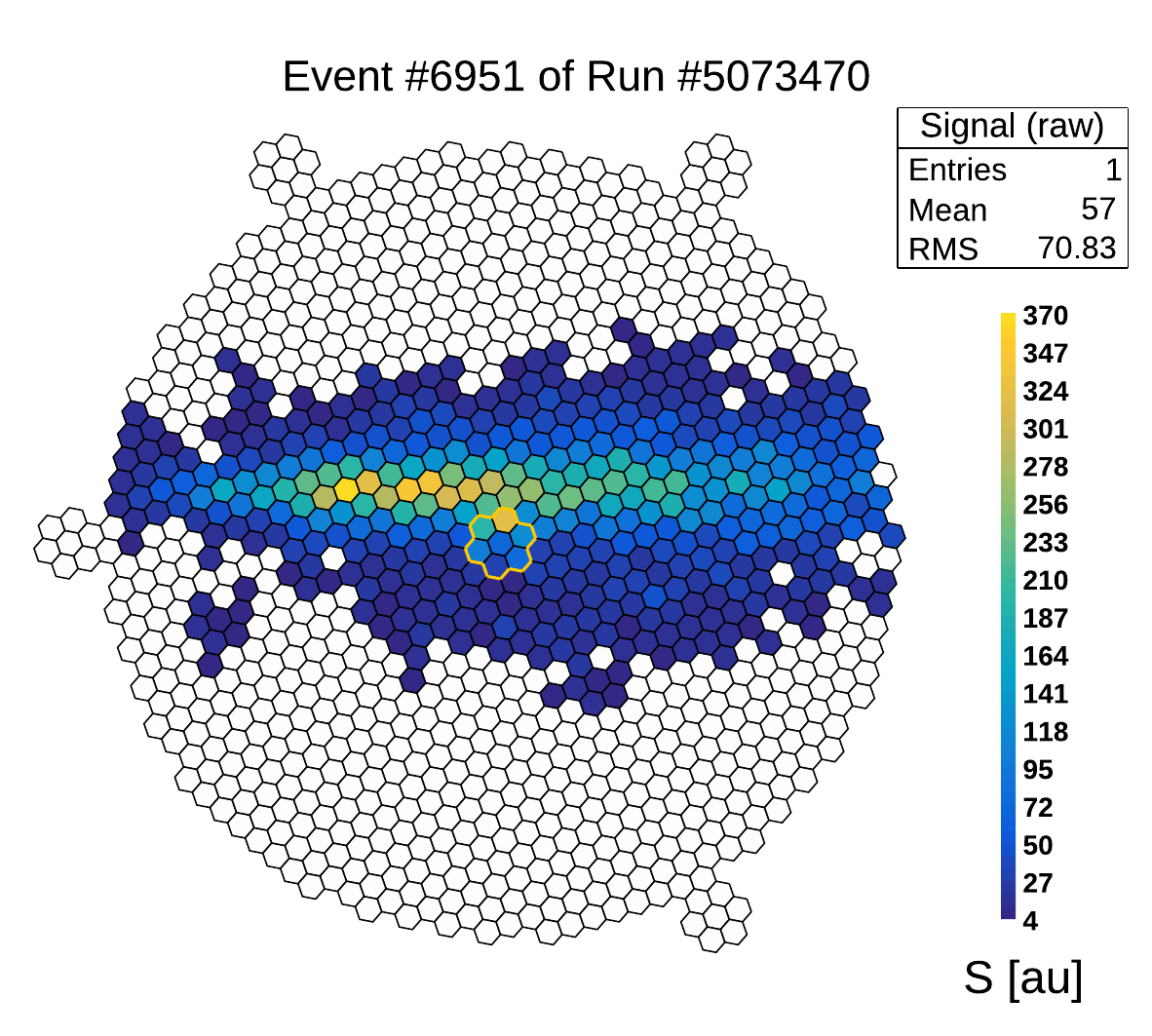}
    \caption{Cherenkov light from extended air shower illuminating a large part of the camera center. The central seven-pixel module is marked by a yellow border.
    }
    \label{fig:camera_image_cherenkov}
\end{figure}

An example of such a large cosmic-ray event is shown in figure \ref{fig:camera_image_cherenkov}. The SiPMs were calibrated using the method described in section \ref{subsec:single_phe_calib}. We cross-checked the obtained calibration with the method described in section \ref{subsec:ffactor} and found good agreement. Comparing the trigger rates of the Hamamatsu SiPMs and neighboring PMTs we found that the SiPMs measure 4.3 times more LoNS. From the shower image analysis, we found that $2.08\pm0.09$ times more Cherenkov photons were detected by the SiPMs.
Selecting only very large events could slightly bias the comparison because these are predominantly produced by high-energy primary protons, which have a lower first interaction point and longer shower development. From such air showers, more UV photons can reach the camera. 
To study this possible effect we produced Cherenkov spectra for proton-induced air showers for nine fixed energies in the range from 100\,GeV to 1\,PeV. We found that while there are differences in the share of UV photons in the Cherenkov light from proton and gamma-ray-induced air showers, the overall responses of the PMT and the SiPM pixels are similar. The maximum difference between MAGIC PMT to Hamamatsu SiPM pixels between the ratios of detected photons from proton air showers with the primary energy between 100\,GeV -- 1\,PeV and gamma-ray air showers of 5\,TeV is 4\,\%. We take this as an additional systematic uncertainty of our measurement. 
These 4\,\% can be considered conservative because the expected number of protons of a high enough energy is very low due to a primary cosmic ray proton spectral index of -2.7 \cite{dampe_collaboration_measurement_2019}.
 Although performing $\sim22\,\%$ better than estimated, this is still in fair agreement with the calculations in section \ref{sec:expectations_spectra}. This is true especially when  considering the systematic uncertainties of the sensors under study as described in section \ref{sec:expectations_spectra} and table \ref{tab:pixel_calc}, the above described effect due to Cherenkov light from proton showers, and the so far unaccounted effects of atmospheric variations \citep{fruck_characterizing_2022}, and LoNS variations at different timescales \cite{roach_characteristic_1958, roach_stable_1963, tarasick_observable_1990, patat_dancing_2008}.

\subsection{Isolated muons Cherenkov light performance}
Isolated muon events from low-impact parameters produce ring or arc shaped images in the camera \cite{vacanti_muon_1994}. Such muon rings can be used for absolute calibration of the light collection efficiency of the IACT \cite{shayduk_calibration_2003, goebel_absolute_2005}. We used the same data set as in section \ref{sec:cherenkov_light_performance} with the SiPM-based module installed in the camera center to reduce the effect of coma aberration that affects pixels towards the camera edge. We found 630 muon ring-shaped events after quality cuts and image cleaning of which 126 produce a signal in the central SiPM module. This number of events is much less than the expected rate of muons because the MAGIC telescopes were operated in stereo-trigger mode. The latter means that a muon event is only saved to disk if there is a simultaneous accidental event measured by the second telescope's imaging camera.

We simulated muon events for the same zenith range as in section \ref{sec:snr}
starting at $285\,\mathrm{g/cm^2}$ which corresponds to an altitude of about 10\,km above sea level.
This is the height at which the muon production from proton air showers reaches its maximum \cite{mirzoyan_tagging_2006}. The additional simulation of muon events is necessary because their Cherenkov light, from several hundred meters above the telescope, gets detected as a ring-shaped image. Due to the short distance, the UV part of such a spectrum is less absorbed in the atmosphere. A dependency on the incident angle of the muon can be neglected, see \cite{gaug_using_2019}.

We measured a mean of $20.5 \pm 0.4$\,phe in the Hamamatsu pixels and $10.8 \pm 0.2$\,phe in the neighboring PMTs. The ratio is $1.91 \pm 0.06_\mathrm{stat}\pm0.22_\mathrm{sys}$, which is in agreement with the simulated value of $1.7$ and shows that our calibration procedure from section \ref{subsec:single_phe_calib} is valid and the nightly uncertainty lays below the aforementioned long-term behavior. The rather large systematic uncertainty was taken from \cite{goebel_absolute_2005} and is extensively discussed in \cite{gaug_using_2019}.

\subsection{Triggering on SiPMs}
During two nights the three SiPM-based modules were connected to the trigger system. We did not physically move the modules in the camera but simply reconnected their analog optical signal output fibers. We performed rate scans to determine the trigger rate as a function of the trigger threshold. These scans were performed during good atmospheric conditions at low zenith distances
observing a source-free region in the night sky. Three PMT modules in symmetric orientation were used for comparison.
Several logic patterns of different compact next neighbor (NN) configurations are available for the trigger setting. A description can be found in \cite{magic_collaboration_major_2016}.

\begin{figure}[htb]
\centering
\includegraphics[width=1\columnwidth]{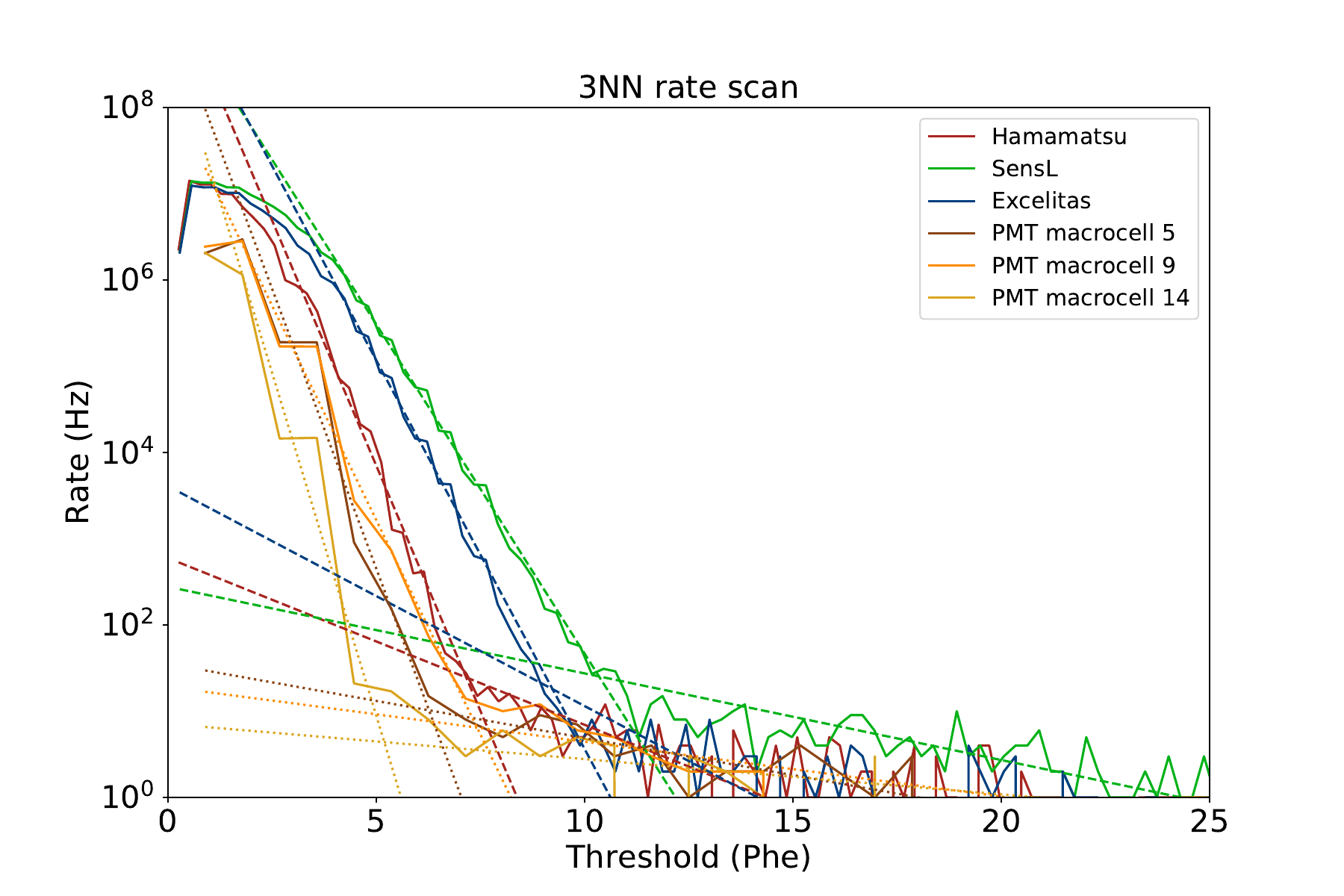}
\caption{3NN rate scan with fitted cosmic-ray and air shower parts.}
\label{fig:3NN_ratescan}
\includegraphics[width=1\columnwidth]{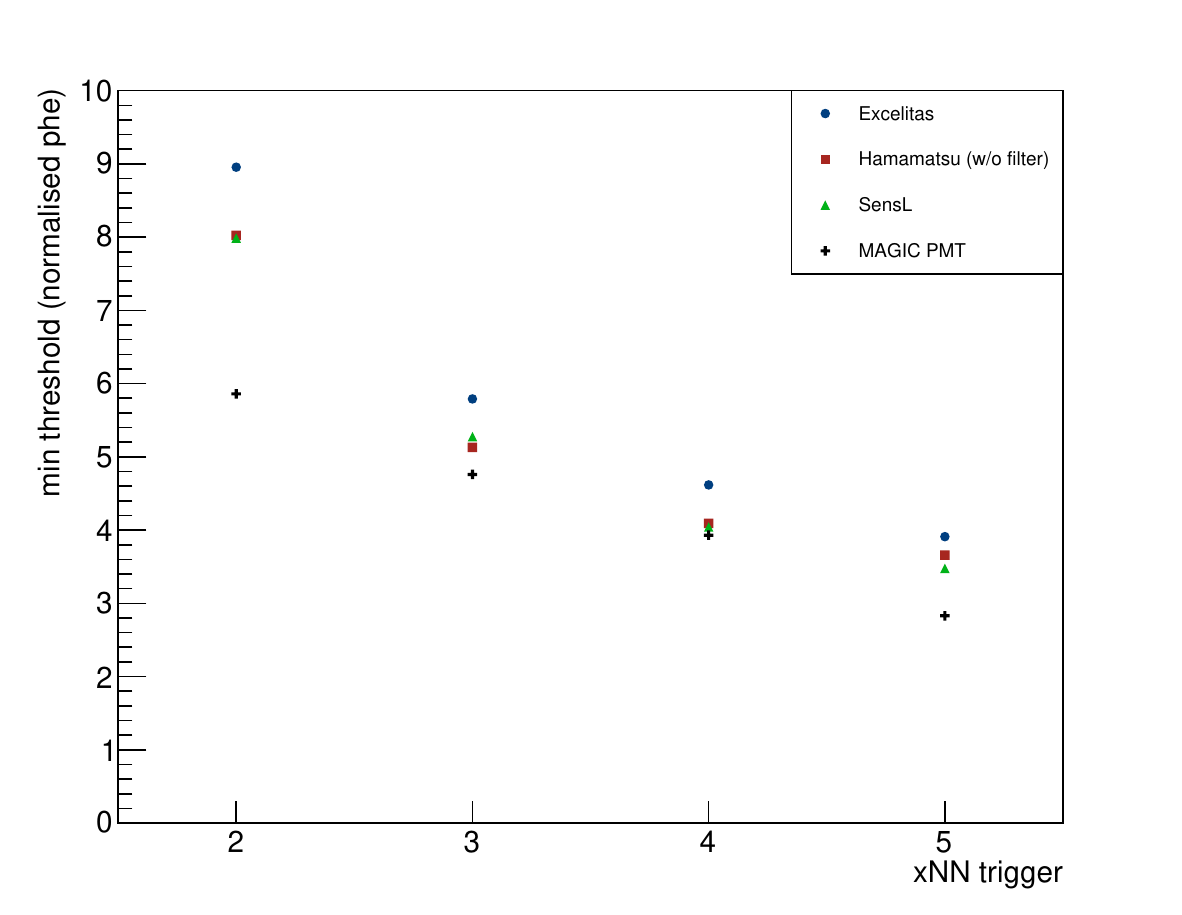}
\caption{Dependence of the threshold on the next neighbor trigger configuration for the sensors under study. 
The threshold value in phe was normalized to the value that the MAGIC PMTs would detect so that the curves can be directly compared with each other.
}
\label{fig:ratescan}
\end{figure}

One rate scan for the 3NN trigger configuration is shown in figure \ref{fig:3NN_ratescan}. The rate scan can be fitted by two distributions, one steep power law due to the LoNS and a more shallow power law due to the cosmic ray events. The crossing of the two distributions defines the minimal trigger threshold for the telescope for the given trigger configuration. The minimal threshold is shown in figure \ref{fig:ratescan} for 2NN, 3NN, 4NN, and 5NN for the three different SiPM module types and the used MAGIC PMTs. It can be seen that the Excelitas, SensL, and Hamamatsu-based SiPM modules result in higher trigger thresholds compared to the PMT-based modules. As for the SNR, it can be seen that SensL, Hamamatsu, and MAGIC PMT performances are  similar, except for the 2NN trigger configuration, whereas the older Excelitas-based module performs worse. For the 3NN trigger configuration, the thresholds can be directly compared to the SNR of section \ref{sec:snr}, which shows agreement within uncertainties.

\begin{figure}[htb]
\centering
\includegraphics[width=1\columnwidth]{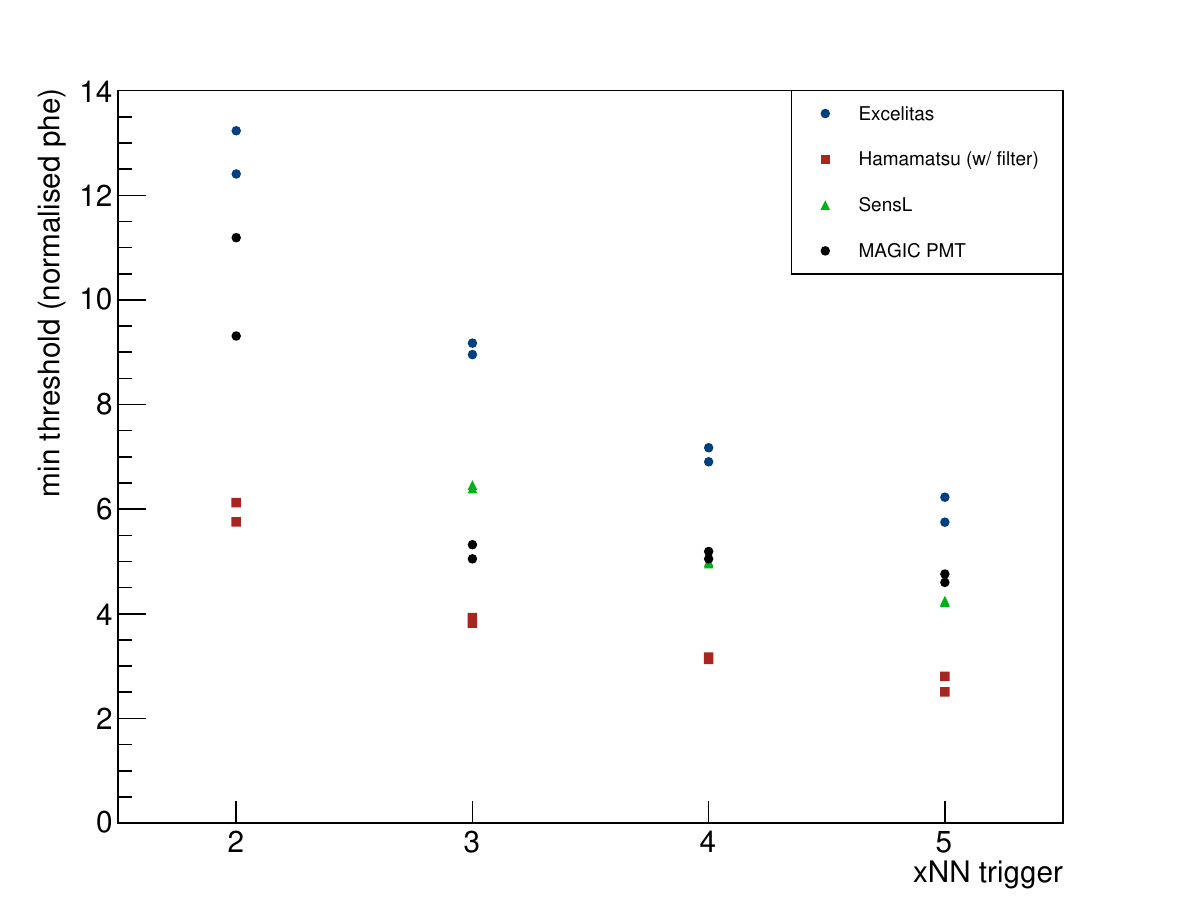}
\caption{Minimal threshold depending on the next-neighbor trigger configuration for the given sensors, same as in figure \ref{fig:ratescan} but with an additional UV-pass filter on top of the Hamamatsu SiPM module. The generally lower thresholds are caused by different atmospheric conditions and a different pointing direction.}
\label{fig:ratescan_hamamatsu_w_filter}
\end{figure}

During another night we mounted a UV-pass filter on top of the Hamamatsu-based SiPM module and took rate scans. The colored glass 335\,nm -- 610\,nm bandpass filter \cite{noauthor_thorlabs_filter} on top of the Hamamatsu SiPM module covered all seven pixels (bandpass filters in general are discussed in section \ref{sec:performance_filter}). We measured the transmission for light incoming perpendicular to the filter and under an angle of 45\degr, which serves as a conservative estimate because this angle is higher than the most extreme case of $31.22^{\circ}$ for light coming reflected from the mirror as discussed in section \ref{sec:acceptance}. The transmissions agree within $^{+4.3\%}_{-0.4\%}$, and the average difference was 1\% in the range 300\,nm -- 700\,nm. Following this, we took the angular filter response for light coming reflected from the mirror as constant.
The result is shown in figure \ref{fig:ratescan_hamamatsu_w_filter}. It can be seen that due to different atmospheric conditions and a different observed star field and therefore LoNS rate, we obtained generally lower thresholds compared to figure \ref{fig:ratescan} but with the same tendencies between the different sensor types. We also note that the trigger rates at the corresponding intersections agreed within uncertainties for both nights. The SiPM modules equipped with Excelitas and SensL SiPMs showed a higher trigger threshold than the PMT modules. In contrast, the Hamamatsu-based pixels with the additional UV bandpass filter provided a lower trigger threshold. 

\begin{figure}[htb]
\includegraphics[width=1\columnwidth]{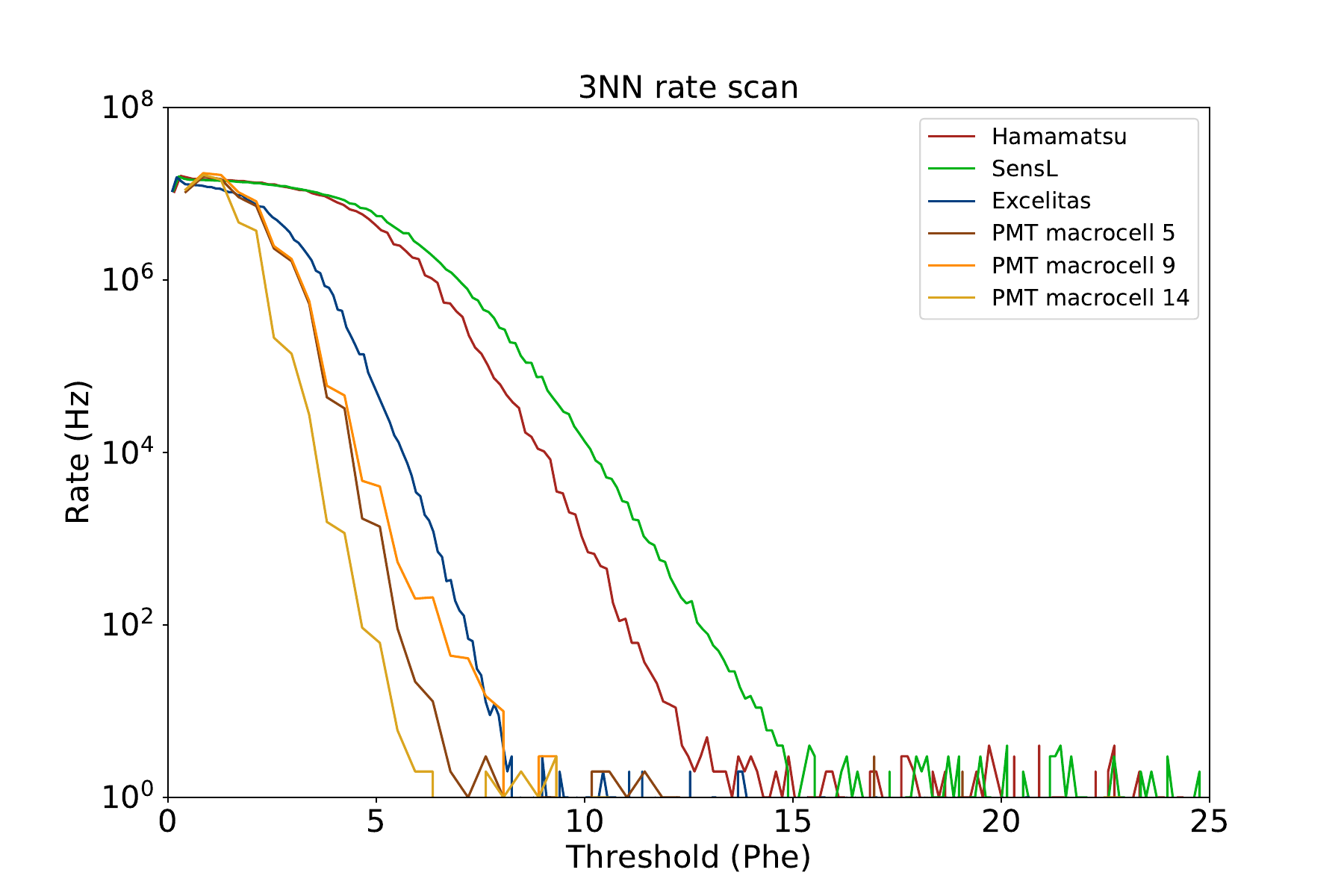}
\caption{3NN rate scan with moderate moonlight. The waning Moon illumination was 70\% and the angular separation to the observed source-free region $\sim50\degr$.}
\label{fig:ratescan_moon}
\end{figure}

In the presence of moonlight the LoNS increases significantly, leading to an elevated trigger threshold. A scan with identical trigger configurations as in Figure \ref{fig:ratescan} but from a different night under the presence of moonlight is shown in Figure \ref{fig:ratescan_moon}. During the time of the measurement, the angular separation between the telescope pointing and the Moon was $\sim50\degr$ and the waning Moon was 70\% illuminated.

A comparison of SiPM and PMT trigger rates can also be found in \cite{hahn_performance_2023}.

We would like to emphasize that the comparison of the thresholds determined by rate scans should not be confused with an equivalent improvement (or lack thereof) in the performance of the entire telescope. Rate scans are only a first step. To understand the telescope threshold for gamma rays, one needs to take into account the trigger, the image reconstruction, its parameterization, and the analysis pipeline; usually, the threshold after the analysis is higher than that just after the trigger \cite{aleksic_major_2016-II}. The study of these is beyond the scope of the current work. However, the SNR of the photosensors is the fundamental parameter for assessing the performance of an imaging camera.

\section{Simulation of bandpass filters}
\label{sec:performance_filter}
To reduce the effect of the LoNS for SiPM-based imaging cameras one can use UV/blue-pass filters (see \cite{montaruli_small_2015,catalano_astri_2018}). Alternatively, the telescope mirrors could be coated by a dielectric coating \citep{forster_dielectric_2013}, the light guides of the pixels could be coated \cite{okumura_development_2023}, or the pixel could be constructed as a so-called Light-Trap, consisting of a SiPM coupled to a polymethylmethacrylate (PMMA) disk doped with a wavelength-shifter \cite{guberman_light-trap:_2019}. The UV/blue pass filter can alternatively be placed directly on the SiPM chips but will increase the crosstalk\cite{mazzillo_electro-optical_2017}. The filter can also be mounted above the surface of the light guides, potentially replacing the "usual" MAGIC protective camera entrance window. In MAGIC, such a setup for attenuating the moonlight was tested several years ago \cite{ahnen_performance_2017}. \cite{arcaro_study_2022} discusses the technical aspects of such a camera window with a dichroic filter and concludes the financial affordability. Also, the MAGIC light detection and ranging (LIDAR) atmospheric monitoring system uses a narrow interference filter for LoNS rejection \cite{fruck_characterizing_2022}. We repeated our LoNS and Cherenkov light simulations from section \ref{sec:snr} with an additional bandpass filter installed in front of the detector modules. For the simulations in this section, we again include the novel PMT R-12992-100, which is used in the imaging cameras of the large (LST) and middle (MST) sized IACTs of CTA \cite{mirzoyan_evaluation_2017}.

\begin{landscape}
\begin{figure*}[p]
\hspace{-6cm}
\includegraphics[height=0.6\textheight]{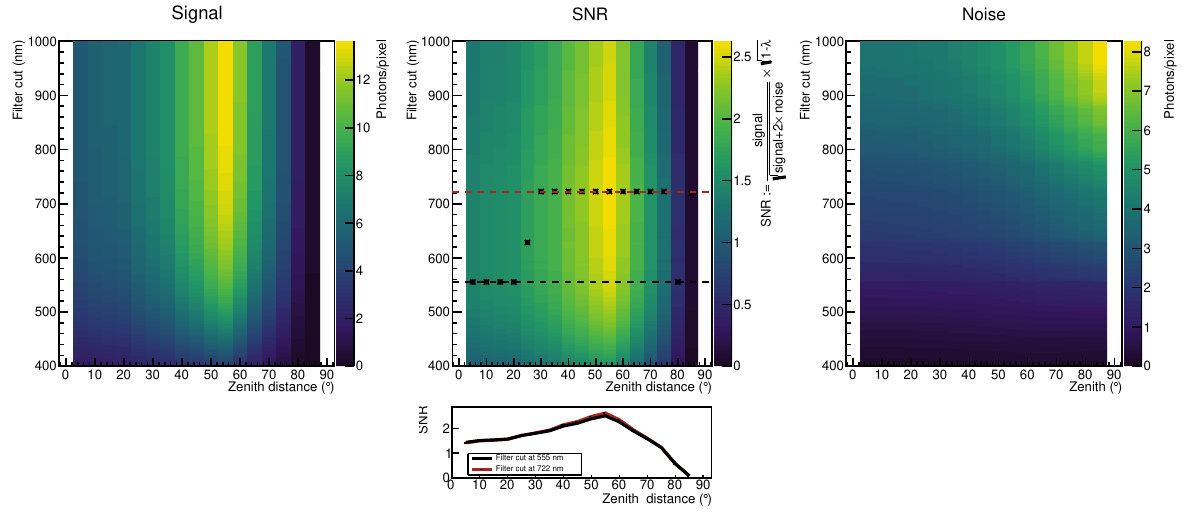}
\caption{2D histograms of Cherenkov photons per pixel (left), LoNS photons per pixel (right), and SNR (center top) in dependence of zenith distance and filter cut-off wavelength. For each simulated zenith distance, the optimal cut-off is marked by a cross in the 2D SNR histogram. The bottom plot shows the SNR for filters with a cut-off wavelength of 555\,nm (black, dashed line) and 722\,nm (red, solid line), also indicated in the center top panel by the dashed lines. Shown are the simulations for the Hamamatsu-based pixels.}
\label{fig:snr_filter_hamamatsu}
\end{figure*}
\end{landscape}

We simulated idealized filters with transmission curves following a Heaviside step function which only allows all photons below a certain cut-off wavelength to pass the filter. We determined the ideal cut-off wavelength in terms of maximizing the SNR as defined in formula \ref{eq:snr_measurement}. An example showing the 2D histograms for the Hamamatsu-based SiPM pixels is presented in figure \ref{fig:snr_filter_hamamatsu}. Depending on the zenith distance, two dominant cut-off wavelengths were found, one at 555\,nm and one at 722\,nm. Comparing to the LoNS spectrum in figure \ref{fig:PDE_LoNS_Cher} one can see that this corresponds to a cut-off just below of the O$\left({}^{1}\mathrm{S}\right)$ peak centered at 557.7\,nm \citep{barth_5577_1961} and just below a set of Meinel OH bands above ${\sim}725$\,nm \citep{meinel_oh_1950}. For few zenith distances, the maximum SNR was achieved with a cut-off wavelength just below the O$\left({}^{1}\mathrm{D}\right)$ line at 630\,nm \citep{roach_stable_1963}. An ideal filter cutting wavelengths of around 550\,nm was also obtained from an MC study of \cite{arcaro_study_2022}. The differences between the three SNRs for filters at 555\,nm, 722\,nm, or 630\,nm are generally very small for SiPM and R-12992-100 pixels and negligible for MAGIC PMT pixels. To exemplify this the SNRs for filters with a cut-off wavelength of 555\,nm and 722\,nm are shown in the center bottom panel of figure \ref{fig:snr_filter_hamamatsu}. The maximal difference in SNR between these two filter cut-offs, which is always reached at a zenith distance of 55\arcdeg, is 6.4\% for the Excelitas SiPMs, 4.5\% for the Hamamatsu SiPMs, 2.9\% for the SensL SiPMs, 0.2\% for the MAGIC PMTs and 0.6\% for the R-12992-100 PMTs.

\begin{figure}[tb]
\centering
\includegraphics[width=1\columnwidth]{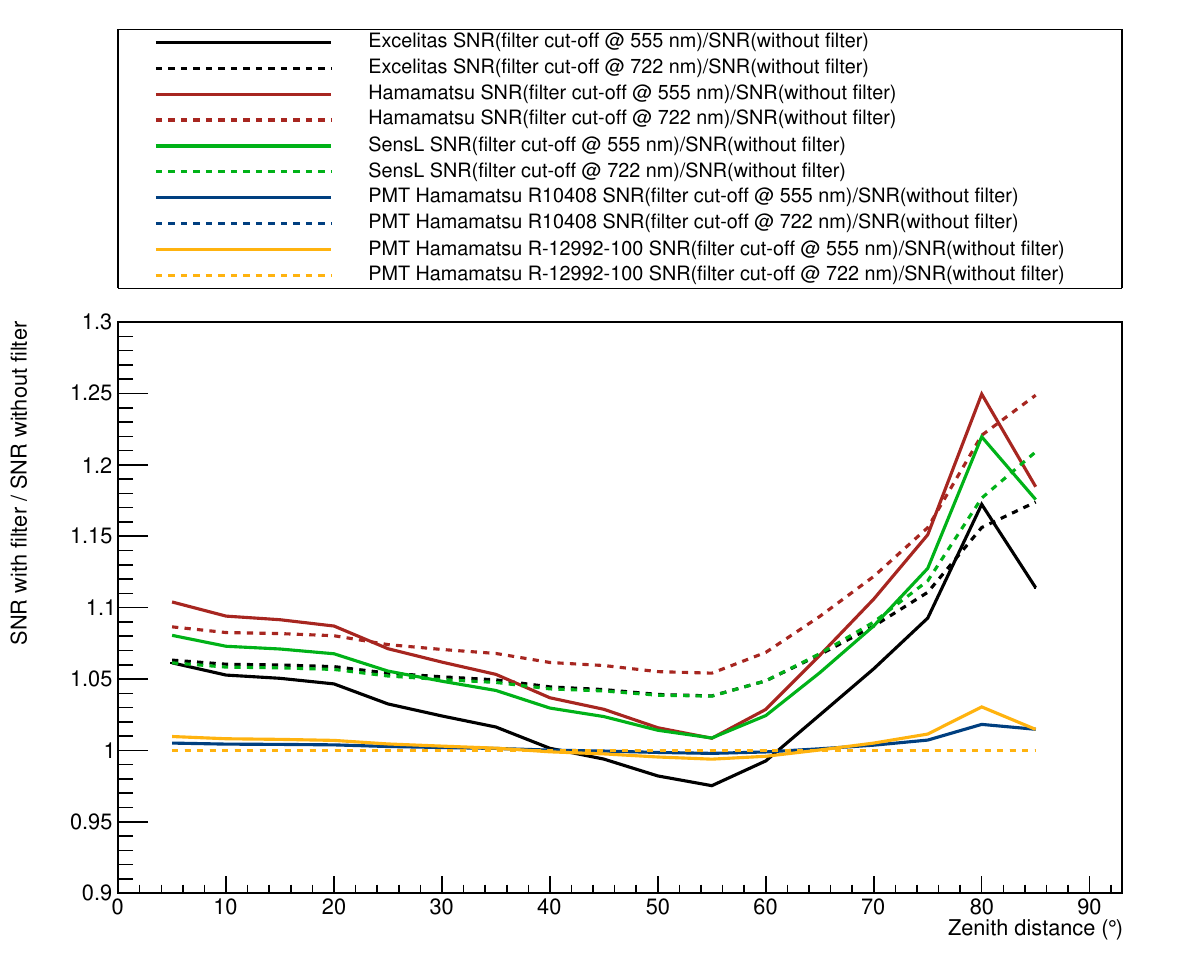}
\caption{Ratios of the SNR of a given light sensor with a given filter cut-off to the same sensor without a filter.}
\label{fig:snr_filter_no_filter_combined}
\end{figure}

\begin{figure}[tb]
\centering
\includegraphics[width=1\columnwidth]{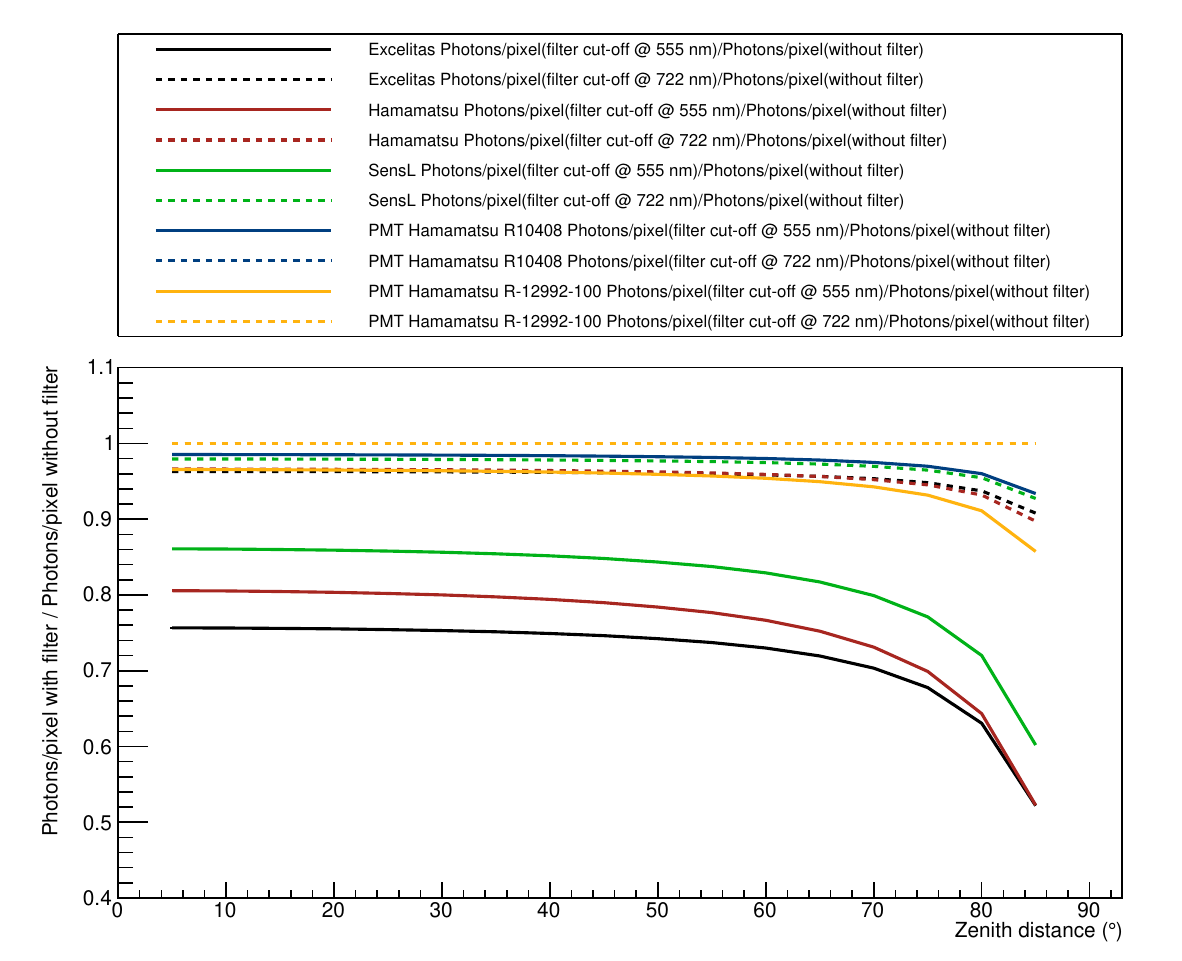}
\caption{Ratios of the Cherenkov photons detected by using filters with given cut-offs to the Cherenkov light without the use of a bandpass filter for the different light sensors.}
\label{fig:signal_filter_no_filter_diff}
\end{figure}

The ratios of the SNR with a filter cut-off at 555\,nm or 722\,nm to the SNR without a filter are shown in figure \ref{fig:snr_filter_no_filter_combined}. Up to 25\% improvement in SNR is possible at very large zenith distances where the LoNS is more intense. One can also see that a filter cut-off at 722\,nm performs much better at medium zenith distances, while there is a smaller difference at small zenith distances. The use of a bandpass filter inevitably also reduces the number of Cherenkov photons. The ratios of the Cherenkov photons detectable by the different pixels are shown in figure \ref{fig:signal_filter_no_filter_diff}. It is obvious that a filter cutoff at 555 nm will dramatically reduce the number of Cherenkov photons detected by SiPM-based pixels. This can seriously deteriorate the telescope's trigger and analysis thresholds, especially at very large zenith distances. Therefore, a cut-off wavelength at 722\,nm seems to offer a reasonable compromise for SiPM pixels as it provides a good SNR improvement and reduces the number of Cherenkov photons by less than 10\% for all zenith distances. The use of such filters for PMTs does not seem to be worth the additional cost and complexity. This is due to the fact that the shape of the wavelength-dependent QE for a PMT with bialkali photocathode makes a very good match with the shape of the Cherenkov light spectrum, see figure \ref{fig:PDE_LoNS_Cher}. Also, in a natural way, it suppresses the most intense part of the LoNS spectrum at longer wavelengths. Therefore, the SNR and the number of Cherenkov photons are influenced much less than for SiPMs.
The performance of an IACT in the first place depends on the discussed SNR, but the final evaluation will be based on the image reconstruction and noise suppression. The detailed study of the influence of a bandpass filter on the trigger threshold and image reconstruction, including the gamma/hadron separation (for MAGIC see \cite{albert_implementation_2008}) is beyond the scope of this work and will be discussed in a dedicated follow-up publication.

\clearpage
\thispagestyle{empty}
\clearpage

\section{Summary}
\label{sec:summary}

The main objective of this study was to compare the performance of SiPM and PMT light sensors in the imaging cameras of IACTs directly and without any assumptions. Although today's SiPMs have comparable or even better peak PDE than the best classical PMTs, due to the strong background from LoNS (noise) and spectral differences between the sensors, this alone is not an indication of their better performance. Only a careful evaluation of the signal-to-noise ratio can provide an answer to the problem under study. 
For this purpose, we built three SiPM-based 7-pixel modules with sensors from three different manufacturers. To achieve a similar active area as for the PMTs in the imaging camera, we developed an analog electronic circuit that actively sums the outputs of seven/nine SiPM chips into a composite pixel, thereby virtually preserving the fast signal timing of a single chip.
We calculated the SNR for SiPM and PMT pixels using simulated LoNS and Cherenkov light spectra and showed that a similar performance can be achieved at most zenith distances with both types of sensors.
At low zenith distances, the PMTs have a slightly higher SNR. At medium to very large zenith distances, our SiPM-based pixels show a comparable SNR to the MAGIC PMTs. The newer PMTs used in the LSTs consistently achieve a higher SNR, while matching our second generation of SiPM-based pixels within uncertainties at ZD$>50\degr$.
It is worth noting that even in the hypothetical absence of the effect of cross-talk, PMTs and SiPMs demonstrate very similar SNRs across all zenith distances, except for marginally higher SNR of SiPMs in the range between $\sim45\degr$ and $\sim65\degr$.
We performed measurements to validate the simulations by using Cherenkov light from cosmic-ray air showers as well as ns fast laser pulses at 355\,nm. Simulations and measurements are in agreement.
We validated our LoNS spectrum model for a wide range of zenith distances by using the signal currents in SiPM and PMT pixels. 
Two types of calibrations were applied, one using the single photoelectron spectrum and the other using the F-factor method.
The measured time resolution showed an improved $T_{0,\mathrm{EAS}}$ which could be of interest for the developing field of intensity interferometry with IACTs \cite{acciari_optical_2020,cortina_first_2022}.
Using rate scans, we showed that a lower trigger threshold can be achieved with SiPMs, but only if a bandpass filter (or some other similar measure) is used to reduce the response to long-wavelength photons in the LoNS spectrum.
Motivated by the increasing number of observations at large zenith distances with MAGIC, we showed the necessity to consider the zenith distance dependence of the Cherenkov and LoNS spectra when selecting a UV/blue-pass filter. This results in two possible (idealized) filters that optimize the SNR for the SiPMs for all zenith ranges.
This study was not targeted at upgrading a specific telescope system. The aim was to underline the importance of the signal-to-noise ratio as the key parameter and provide useful information. So, unlike \cite{arcaro_study_2022}, we refrained from simulating a full telescope sensitivity for a specific sensor type, analysis pipeline, and (standard candle) source spectrum.

\section*{Acknowledgements}
We are very grateful to the members of the MAGIC collaboration for allowing us to install the SiPM modules in the MAGIC-I imaging camera and operate them routinely for many years.
We would like to express our gratitude to the electronics and mechanical engineers at the Max Planck Institute for Physics (MPP) Werner Haberer, Ronald Maier, and Holger Wetteskind.
We are thankful to the local La Palma crew Javier Herrera, Victor Acciari, Auni Somero, Marie Karjalainen, and Eduardo Colombo for their extensive support, and we thank Julian Sitarek for his fruitful inputs to this manuscript.
Thanks should also go to the shift crews helping with the data taking, especially to Martin Will (MPP) and Juliane van Scherpenberg (MPP).
We would like to thank the Instituto de Astrofísica de Canarias (IAC) for the excellent working conditions at the Observatorio del Roque de los Muchachos in La Palma.
We acknowledge the support and expertise in the field of LoNS and optical spectroscopy by Chris Benn (Isaac Newton Group of Telescopes (INT)), Cecilia Farina (INT), Clár-Brid Tohill (INT) and Josefa Becerra (IAC).
This work was also supported by Dieter Renker (Technical University Munich) and Eckart Lorenz (MPP) with their expertise during the early phases of the project.
We thank the Max Planck Institute for Physics for their excellent technical support and the companies Hamamatsu, SensL, and Excelitas for their fruitful cooperation.

\bibliography{SiPM_paper_bibfile}

\end{document}